\documentclass[twocolumn]{aastex63}

\usepackage{lineno}
\usepackage{amsmath}
\usepackage{amssymb}
\usepackage{graphicx}
\usepackage{color}
\usepackage{mathrsfs}
\usepackage{epsfig}
\usepackage{comment}
\usepackage{ulem}
\usepackage[T1]{fontenc}

\usepackage[graphicx]{realboxes}
\usepackage{booktabs,array}

\newcommand{\msun}{{\rm M}_\odot} 
\newcommand{\zsun}{{\rm Z}_\odot} 

\newcommand{\mbh}{M_{\rm BH}} 
\newcommand{\tinf}{T_\infty} 
\newcommand{\cinf}{c_{\rm s,\infty}} 
\newcommand{\mach}{\mathcal{M}} 

\newcommand{\rb}{R_{\rm B}} 

\newcommand{\rmin}{r_{\rm min}} 
\newcommand{\rmax}{r_{\rm max}} 

\newcommand{\ledd}{L_{\rm E}} 
\newcommand{\medd}{\dot{M}_{\rm E}} 
\newcommand{\ledduv}{L_{\rm E,UV}} 
\newcommand{\medduv}{\dot{M}_{\rm E,UV}} 
\newcommand{\leddir}{L_{\rm E,IR}} 
\newcommand{\meddir}{\dot{M}_{\rm E,IR}} 

\newcommand{\mdot}{\dot{M}_{\rm acc}} 
\newcommand{\mcrit}{\dot{M}_{\rm crit}} 

\newcommand{\mdott}{\langle \dot{M}_{\rm acc}\rangle} 

\newcommand{\minf}{\dot{M}_{\rm in}} 
\newcommand{\fin}{F_{\rm in}} 
\newcommand{\linf}{l_{\rm in}} 
\newcommand{\fk}{F_{\rm K}} 
\newcommand{\rcent}{R_{\rm cent}} 
\newcommand{\rioni}{R_{\rm HII}} 
\newcommand{\rsb}{R_{\rm sb}} 

\newcommand{\mout}{\dot{M}_{\rm out}} 
\newcommand{\moutt}{\langle \dot{M}_{\rm out}\rangle } 

\newcommand{\mh}{M_{\rm h}} 

\received{\today}

\shorttitle{Super-Eddington mass growth of IMBHs}
\shortauthors{Toyouchi et al.}

\begin{document}

\title{Super-Eddington mass growth of intermediate-mass black holes embedded in dusty circumnuclear disks}

\correspondingauthor{Daisuke Toyouchi, Kohei Inayoshi}
\email{d.toyouchi@gmail.com, inayoshi.pku@gmail.com}

\author[0000-0003-3467-6079]{Daisuke~Toyouchi}
\affil{Kavli Institute for Astronomy and Astrophysics at Peking University, Beijing 100871, China}
\affil{Kavli Institute for the Physics and Mathematics of the Universe (Kavli IPMU, WPI), The University of Tokyo, Chiba 277-8583, Japan}
\affil{Theoretical Astrophysics Group, Department of Physics, Kyoto University, Sakyo-ku, Kyoto 606-8502, Japan}

\author[0000-0001-9840-4959]{Kohei Inayoshi}
\affil{Kavli Institute for Astronomy and Astrophysics, Peking University, Beijing 100871, China}

\author[0000-0003-3127-5982]{Takashi~Hosokawa}
\affil{Theoretical Astrophysics Group, Department of Physics, Kyoto University, Sakyo-ku, Kyoto 606-8502, Japan}

\author[0000-0003-2309-8963]{Rolf~Kuiper}
\affil{Institute of Astronomy and Astrophysics, University of T\"ubingen, Auf der Morgenstelle 10, D-72076 T\"ubingen, Germany}

\begin{abstract}

We perform the first three-dimensional radiation hydrodynamical simulations that investigate 
the growth of intermediate-mass BHs (IMBHs) 
embedded in massive self-gravitating, dusty nuclear accretion disks.
We explore the dependence of mass accretion efficiency on the gas metallicity $Z$ and
mass injection at super-Eddington accretion rates from the outer galactic disk $\minf$, and find that
the central BH can be fed at rates exceeding the Eddington rate only when
the dusty disk becomes sufficiently optically thick to ionizing radiation.
In this case, mass outflows from the disk owing to photoevaporation is suppressed and thus
a large fraction ($\gtrsim 40\%$) of the mass injection rate can feed the central BH.
The conditions are expressed as $\minf > 2.2\times 10^{-1}~\msun ~{\rm yr}^{-1}
(1+Z/10^{-2}~\zsun)^{-1}(c_{\rm s}/10~{\rm km~s}^{-1})$,
where $c_{\rm s}$ is the sound speed in the gaseous disk.
With increasing numerical resolution, 
vigorous disk fragmentation reduces the disk surface density and dynamical heating by 
formed clumps makes the disk thickness higher.
As a result, the photoevaorative mass-loss rate rises and thus the critical injection rate increases for fixed metallicity.
This process enables super-Eddington growth of BHs until the BH mass reaches $\mbh \sim 10^{7-8}~\msun$,
depending on the properties of the host dark-matter halo and metal-enrichment history.
In the assembly of protogalaxies, seed BHs that form in overdense regions with 
a mass variance of 3-4$\sigma$ at $z\sim 15-20$ are able to undergo short periods of their rapid growth 
and transits into the Eddington-limited growth phase afterwards to be supermassive BHs observed at $z>6-7$.

\end{abstract}


\section{INTRODUCTION} \label{sec:intro}

The formation process of supermassive black holes (SMBHs) is one of the most important 
puzzles in modern astrophysics.
The existence of SMBHs with $\mbh \gtrsim 10^9~\msun$ in the early universe ($z \gtrsim 6-7$) provides a stringent constraint for their mass-growth timescale \citep[e.g.,][]{Fan2001, Willott2010, Mortlock2011, Venemans2013, Wu2015, Banados2018, Matsuoka2019, Onoue2019, Yang2020}.
Various models for their seed black holes (BHs) have been suggested \citep[see e.g.,][for a review]{Volonteri2012, Haiman2013, Inayoshi2019}. 
A natural candidate is Pop III remnant BHs with a typical mass of $\mbh \sim 10^2 \ M_\odot$ \citep[e.g.,][]{Yoshida2008, Hosokawa2011, Hosokawa2016, Susa2014, Hirano2014, Hirano2015, Stacy2016, Sugimura2020b}.
In this case, they must undergo substantially high accretion rates exceeding the Eddington limit to reach $\mbh \sim 10^9~\msun$ by $z \sim 7$.
Another possibility is producing more massive seed BHs with $\mbh \sim 10^{4\mathchar`-5} \ M_\odot$ via the direct collapse of massive pristine gas through formation of supermassive stars \citep[e.g.,][]{Omukai2001, Oh2002, Bromm2003, Hosokawa2012, Inayoshi2012, Inayoshi2014a, Regan2014, Visbal2014, Sugimura2014, Sugimura2016, Latif2016, Umeda2016, Chon2016, Chon2018, Hirano2017, Wise2019} or runaway stellar mergers in dense clusters \citep[e.q.,][]{Omukai2008, Devecchi2009, Katz2015, Tagawa2015, Tagawa2020, Yajima2016, Sakurai2017, Sakurai2019}.
With a head start in mass, the Eddington-limited accretion allows seed BHs to grow up to 
SMBHs by $z\gtrsim 6$, but a high duty cycle of {\it O}(1) is still required.
Therefore, in any seeding models, it is essential whether rapid growth of BHs could be 
sustained continuously in protogalaxies.

Many theoretical and numerical studies have confirmed that super-Eddington accretion flows 
are feasible inside the photon trapping radius, where radiation is advected with accreting matter 
before escaping via diffusion \citep[e.g.,][]{Abramowicz1988, Watarai2000, Ohsuga2005, Ohsuga2011, Jiang2014, Yang2014, Yang2018, Sadowski2016}.
Radiation hydrodynamics (RHD) simulations that cover the BH gravitational sphere of influence 
showed that the mass accretion rate is generally self-regulated below the Eddington value 
due to the outward thermal pressure gradient induced via photoionization and heating 
\citep[e.g.,][]{Milosavljevic2009a, Milosavljevic2009b, Park2011, Park2012, Jeon2012, Park2017b}. 
Accordingly, several possible scenarios of super-Eddington accretion from larger scales 
($\sim 1-10$ pc) have been suggested.
\cite{Inayoshi2016} showed that when a BH is embedded in sufficiently dense gas with
a density of $n_{\rm H} \gtrsim 10^5 \ {\rm cm}^{-3} (\mbh /10^4 \ M_\odot)^{-1}$, 
photoionization / heating of gas is suppressed due to efficient recombination,
leading to rapid mass accretion onto the BH without being impeded by radiative feedback
\citep[see also][]{Sakurai2016, Park2016, Park2020}.
In the intense inflow, the inward ram pressure of accreting gas substantially overcomes the 
sum of outward thermal pressure and radiation force.
They also found that such a dense environment would be realized in the nuclei of high-$z$ protogalaxies without prior star formation and seed BHs that migrate to the region within 
a Hubble timescale would rapidly grow into $\mbh \gtrsim 10^5~\msun$ at 
hyper-Eddington rates ($\gg \medd$).
Moreover, anisotropic radiation emitted from the nuclear disk toward the polar regions
dramatically reduces the negative feedback effect because gas accretion is allowed through
the equatorial region that is shielded against intense ionizing radiation from the accreting BH
\citep{Sugimura2017,Takeo2018}.
Mechanical feedback due to strong outflows launched from the disk completely evacuate the polar regions
but does not affect the gas dynamics.
In fact, even if a significant fraction of mass is loaded into outflows, the emergent radiation becomes less intense
and thus the super-Eddington accretion rate through the disk still holds \citep{Takeo2020}.

Although those scenarios are potentially intriguing, several simplified treatments are still imposed in their RHD simulations.
One of them is the absence of angular momentum of accreting gas that is supplied from larger galactic scales.
\cite{Sugimura2018} studied the effect of angular momentum of gas and showed that 
mass accretion of rotating gas is suppressed from the standard Bondi rate when 
the centrifugal radius is larger than the BH gravitational influence radius and the angular momentum transport is inefficient.
Therefore, it is crucial to quantify the efficiency of angular momentum redistribution owing to 
gravitational torques caused by spiral arms or turbulent motions excited within the circum-nuclear disk.
In the previous studies, primordial chemical composition of gas is commonly assumed.
However, recent observations have reported that the nuclear regions and host galaxies of bright quasars at $z \gtrsim 6$ 
already contain a large amount of dust \citep[e.g.,][]{Venemans2012, Venemans2017}.
The existence of heavy elements generally affect the thermal properties of gas and thus could change the mass growth of SMBHs.  
A series of 1D RHD simulations \citep{Yajima2017a, Toyouchi2019} have investigated accretion of dusty gas onto BHs and found that 
radiative force upon high-opacity dusty accreting flows strongly regulates mass accretion onto the central BHs, 
so that super-Eddington flows are prohibited especially for $Z \gtrsim 10^{-2}~\zsun$.
In this paper, we extend our previous study and perform 3D RHD simulations adopting more realistic configuration 
of dusty and rotating accretion flows.

In particular, we explore the accretion dynamics at physical scales of $\sim$ 0.01-1 pc, which roughly corresponds to the size of 
dusty tori or circum-nuclear disks (CNDs) that are expected to play an essential role in fueling the central active galactic nuclei 
(AGNs) \citep[e.g.,][]{Hicks2013, Izumi2016}.
The dynamics of dusty nuclear disks has been extensively studied with a series of 3D hydrodynamical simulations 
\citep{Wada2002, Wada2009, Wada2016, Wada2018} that successfully reproduce the observed spectral features of
gaseous structure in low-luminosity AGNs \citep{Izumi2018}.
However, they focus on the sub-Eddington AGN population in the local universe, considering already grown massive BHs
with $\mbh \gtrsim 10^7~\msun$ and metal-enriched accretion disks with the solar abundance composition.
We here focus on intermediate mass BHs (IMBHs) with $\mbh = 10^4~\msun$ embedded in low-metallicity environments
($Z = 10^{-3}\mathchar`-10^{-1}~\zsun$) to study super-Eddington mass growth of seed BHs in the early universe.

The rest of the paper is organized as follows.
We first describe the numerical method and settings of our 3D RHD simulations in Section \ref{sec:method}. 
The main results of our numerical simulations and a theoretical explanation for them are given in Section \ref{sec:result} and \ref{sec:condition}, respectively.
Based on these results, we further argue whether super-Eddington accretion can happen in the early universe in Section \ref{sec:possible}. 
Additionally, we provide discussions regarding physical processes which are not incorporated in our current simulations in Section \ref{sec:effect}. 
Finally, the summary and conclusion are given in Section \ref{sec:summary}.


\section{SIMULATION METHOD} \label{sec:method}

We utilize a hydrodynamical simulation code \citep[{\tt PLUTO} 4.1; ][]{Mignone2007}, 
which has been modified to study massive star formation and evolution of proto-planetary disks 
\citep[e.g.,][]{Kuiper2010,Hosokawa2016,Nakatani2018a, Nakatani2018b, Kuiper2018, Kolligan2018, Nakatani2019, Fukushima2020}. 
In particular, we make use of the specific version of the code adjusted for investigating BH accretion physics under radiative feedback 
\citep[][]{Sugimura2017, Sugimura2018, Toyouchi2019,Toyouchi2020}.


\subsection{Basic Equations} \label{sec:basic}

We here perform three-dimensional hydrodynamical simulations to investigate the accretion dynamics of a gaseous disk 
surrounding a nuclear BH, which is located at the origin of spherical coordinates of ($r$, $\theta$, $\phi$).
The basic equations of hydrodynamics that we solve are the following:
the equation of continuity,
\begin{eqnarray}
\frac{\partial \rho}{\partial t} + \nabla \cdot (\rho \bm{v}) = 0,  
\label{eq:mass_cons}
\end{eqnarray}
and the equations of motion,
\begin{eqnarray}
\frac{\partial \rho v_r}{\partial t} + \nabla \cdot (\rho v_r \bm{v}) = - \frac{\partial P}{\partial r} + \rho \frac{v^2_\theta + v^2_\phi}{r} + \rho g_r \ ,  
\label{eq:mom_cons_r}
\end{eqnarray}
\begin{equation}
\begin{split}
\frac{\partial \rho v_\theta}{\partial t} + \nabla \cdot (\rho v_\theta \bm{v}) = &- \frac{1}{r}\frac{\partial P}{\partial \theta} - \rho \frac{v_\theta v_r}{r} \\
&+ \rho \frac{v^2_\phi~{\rm cot}~\theta}{r} + \rho g_\theta \ ,  
\label{eq:mom_cons_t}
\end{split}
\end{equation}
\begin{equation}
\begin{split}
\frac{\partial \rho v_\phi}{\partial t} + \nabla \cdot (\rho v_\phi \bm{v}) = &- \frac{1}{r~{\rm sin}~\theta}\frac{\partial P}{\partial \phi} - \rho \frac{v_\phi v_r}{r} \\
&- \rho \frac{v_\phi v_\theta~{\rm cot}~\theta}{r} + \rho g_\phi \ ,
\label{eq:mom_cons_p}
\end{split}
\end{equation}
%
where $\rho$, is the gas density, $\bm{v} = (v_r, v_\theta, v_\phi)$ is the velocity vector, $P$ is the gas pressure, 
$\bm{g} = (g_r, g_\theta, g_\phi)$ characterizes the external force due to the BH gravity, gas self-gravity and absorption and scattering of radiation.

We solve the energy equation of
\begin{eqnarray}
\frac{\partial E}{\partial t} + \nabla \cdot (H \bm{v}) = \rho~\bm{v} \cdot \bm{g} + \rho~(\Gamma - \Lambda),  
\label{eq:ene_cons}
\end{eqnarray}
where $E$ is the total (internal and kinetic) energy density, $H$ is the enthalpy per unit volume, 
and $\Gamma$ and $\Lambda$ the specific heating and cooling rates in units of erg s$^{-1}$ g$^{-1}$.
We set a minimum temperature floor of 100 K and turn gas cooling off when the (local) Jeans length becomes unresolved 
with the longest size of each grid cell.
Stellar feedback and star formation within a gravitationally unstable disk are not considered in this study, 
but the potential importance on BH growth is discussed in Section \ref{sec:stellar}.

We estimate the heating and cooling rates by solving a chemical reaction network of metal-polluted gas, 
which is composed of the following eight species of HI, HII, HeI, HeII, HeIII, CII, OI, and e$^-$.
The number density of the $i$-th species $n_i$ is calculated with the non-equilibrium rate equation of
\begin{eqnarray}
\frac{\partial n_i}{\partial t} + \nabla \cdot (n_i \bm{v}) = n_{\rm H} R_i,  
\label{eq:chem_cons}
\end{eqnarray}
where $R_i$ is the sum of the reaction rate coefficients related to the $i$-th composition and $n_{\rm H}$ is the number density of hydrogen nuclei.
The CII and OI abundances are set to $n_{\rm CII}/n_{\rm H} = 0.927 \times 10^{-4}~Z/Z_\odot$ and $n_{\rm OI}/n_{\rm H} = 3.568 \times 10^{-4}~Z/Z_\odot$.
We here consider dust grain in metal-polluted gas, assuming that the dynamics of dust perfectly follows hydrodynamics so that a constant 
dust-to-gas mass ratio of $0.01~Z/Z_\odot$ is kept.
We take into account 9 reactions including photoionization and collisional ionization of HI, HeI and HeII, and recombination of HII, HeII, HeIII.
With the updated chemical abundances, we compute $\Lambda$ and $\Gamma$ summing up the contributions of photoelectric heating, 
fine-structure lines of CII and OI, free-free emission of HI, HeI and HeII, and dust-gas collisional energy transfer.
Other heating and cooling processes via heavy elements, e.g., meta-stable line cooling and photoionization heating, hardly affect the thermal properties 
of low-metallicity gas with $Z \leq 0.1~Z_\odot$ considered in our study \citep[see also][]{Dere2009, Milosavljevic2009b, Draine2011}.
In addition, we neglect molecular components such as H$_2$ and CO in the low-metallicity gas
because the timescale of their formation on the surface of dust grains is generally much longer than 
the dynamical timescale of a gaseous disk \citep{Krumholz2012}.

Additionally, we also consider FUV and X-ray background radiation as heating sources.
Supposing an early galaxy formation phase with active star formation, we assume the background fields 100 times stronger than in the solar neighborhood \citep[see][for more detailes]{Toyouchi2019}.
With the background heating, the equilibrium gas temperature is maintained above our floor value at gas number density of $\lesssim 10^{6}~\rm cm^{-3}$, which is applicable to most grids except for inner disk parts in some of our simulations.


\begin{table*}
\begin{center}
\caption{Model parameters and results} \label{table:model}
\begin{tabular}{ccccccccc} \hline \hline
Model                & $Z \ [{\rm Z_\odot}]$  & $\fin$        & UV rad. force         & IR rad. force          & Equatorial sym.         & $(N_r, N_\theta, N_\phi)$ & $\mdott/\minf$ & $\moutt/\minf$ \\ \hline
Z-2F2                  & $10^{-2}$                  & $100$     & YES                       & YES                      & YES                           & (128, 36, 72)                     & 0.38            & 0.37 \\
Z-2F1                  & $10^{-2}$                  & $10$       & YES                       & YES                      & YES                           & (128, 36, 72)                     & 0.001          & 0.91 \\
Z-2F3                  & $10^{-2}$                  & $1000$   & YES                       & YES                      & YES                           & (128, 36, 72)                     & 0.75            & 0.23 \\
Z-1F2                  & $10^{-1}$                  & $100$     & YES                       & YES                      & YES                           & (128, 36, 72)                     & 0.58            & 0.26 \\
Z-1F1                  & $10^{-1}$                  & $10$       & YES                       & YES                      & YES                           & (128, 36, 72)                     & 0.11            & 0.67 \\
Z-3F2                  & $10^{-3}$                  & $100$     & YES                       & YES                      & YES                           & (128, 36, 72)                     & 0.02            & 0.96 \\
Z-3F3                  & $10^{-3}$                  & $1000$   & YES                       & YES                      & YES                           & (128, 36, 72)                     & 0.42            & 0.51 \\ \hline
Z-2F2nuv            & $10^{-2}$                  & $100$     & NO                         & YES                      & YES                           & (128, 36, 72)                     & 0.39            & 0.39 \\
Z-2F2nir              & $10^{-2}$                  & $100$     & YES                       & NO                        & YES                           & (128, 36, 72)                     & 0.37            & 0.36 \\
Z-2F2hr               & $10^{-2}$                  & $100$     & YES                       & YES                      & YES                           & (128, 36, 144)                   & 0.25            & 0.38 \\ 
Z-2F2ne              & $10^{-2}$                  & $100$     & YES                       & YES                      & NO                             & (128, 72, 72)                     & 0.14            & 0.72 \\ 
Z-2F2hr+ne         & $10^{-2}$                  & $100$     & YES                       & YES                      & NO                             & (128, 72, 144)                   & 0.004          & 0.93 \\ 
Z-2F3hr+ne         & $10^{-2}$                  & $1000$   & YES                       & YES                      & NO                             & (128, 72, 144)                   & 0.02            & 0.74 \\  \hline
\end{tabular}
\end{center}
\end{table*}

\subsection{Radiative Feedback} \label{sec:subgrid}

Our simulation adopts a subgrid model to incorporate the radiative feedback against the accretion flow onto BHs.
We suppose that the inside of $\rmin$ is a sink region, in which a circum-BH accretion disk is contained.
Mass accretion rates onto the unresolved disk $\mdot$ are evaluated with the inward mass flux measured at $\rmin$ at each time step.
Photons produced via the mass accretion are injected from the sink, and the luminosity $L$ is described with the fitting formula given by \cite{Watarai2000}
\begin{eqnarray}
L =  
\begin{cases}
2 \ \ledd \ \left [ 1 + {\rm ln} \left ( \frac{\dot{M}}{2 \medd} \right ) \right ] & (\dot{M} > 2 \medd) \\
\ledd \ \frac{\dot{M}}{\medd} & ({\rm otherwise}) \ \ ,
\end{cases}
\label{eq:MtoL}
\end{eqnarray}
where $\ledd$ and $\medd$ are the Eddington luminosity and mass accretion rate defined as below, 
\begin{eqnarray}
\ledd = \frac{4 \pi G \mbh c}{\kappa_{\rm T}} = 3.3 \times 10^8~L_\odot \left ( \frac{\mbh}{10^4 \ {\rm M_\odot}} \right ) \ ,
\label{eq:ledd}
\end{eqnarray}
\begin{eqnarray}
\medd = \frac{\ledd}{\eta c^2} = 2.2 \times 10^{-4}~M_\odot \ {\rm yr}^{-1} \left ( \frac{\mbh}{10^4 \ {\rm M_\odot}} \right ) \ ,
\label{eq:medd}
\end{eqnarray}
where the opacity of Thomson scattering is $\kappa_{\rm T} = 0.4~{\rm cm^2~g^{-1}}$, and the radiative efficiency is assumed $\eta = 0.1$. 
This formula implies that the luminosity $L$ does not greatly exceeds $\ledd$ owing to the photon trapping effect even in $\mdot > \medd$.

In this study, we assume an anisotropic radiation from the unresolved circum-BH disk,
\begin{eqnarray}
F(\theta) =  \frac{L}{4\pi r^2} f(\theta) \ , \ \ f(\theta) \propto {\rm cos}^2 \theta \ ,
\label{eq:aniso_flux}
\end{eqnarray}
where the anisotropic factor is normalized according to $\int f(\theta) {\rm d}\Omega = 4 \pi$.
This anisotropic radiation field is based on the RHD simulation of super-Eddington accretion flow by \cite{Ohsuga2005}, suggesting that photons are preferentially emitted perpendicular to the accretion disk plane.

We consider a power-law spectrum $L_\nu \propto \nu^{-\alpha} \ (\alpha = 1.5)$ in a UV range $6~{\rm eV} \leq h \nu \leq 1~{\rm keV}$ and solve the frequency-dependent radiative transfer along the radial cells.
We take into account the consumption of photons by photoionization of HI, HeI, and HeII with the cross-sections given by \cite{Osterbrock1989} and \cite{Yan1998}, and the dust attenuation with the opacity table of \cite{Weingartner2001}. 
In addition to the UV photons injected from the sink, we also solve the transfer of diffuse IR photons coming from the thermal dust emission with the flux-limited diffusion (FLD) approximation method. 
We utilize the FLD module developed by \cite{Kuiper2010, Kuiper2020}, which has been applied to a lot of studies especially for the formation of massive stars \citep[e.g.,][]{Kuiper2011, Kuiper2012, Kuiper2018}.

Our simulation takes into account the radiative force via Thomson scattering, photoionization, and dust absorption consistently with the obtained radiation fields. 
It is worth noting that the dust absorption of UV photon dominates the radiative force for $Z \gtrsim 10^{-3}~\zsun$ and effectively modifies the Eddington limit as below,
\begin{eqnarray}
\ledduv &=& X_{\rm d,UV} \ledd \ ,
\label{eq:ledduv}
\end{eqnarray}
\begin{eqnarray}
\medduv &=& \frac{\ledduv}{\eta c^2} \ ,
\label{eq:medduv}
\end{eqnarray}
\begin{equation}
X_{\rm d,UV} \equiv \frac{\kappa_{\rm T}}{\kappa_{\rm T} + \kappa_{\rm d,UV}} = \left \{ {1 + 7.1 \left (\frac{Z}{10^{-2} \ Z_\odot} \right )} \right \}^{-1} \ ,
\label{eq:kaiuv}
\end{equation}
where $\kappa_{\rm d,UV}$ is the the dust absorption opacity for UV photon, for which we assume $\kappa_{\rm d,UV} = 2.8 \times 10^2 (Z/Z_\odot)~{\rm cm^2~g^{-1}}$ \citep[e.g.,][]{Yajima2017a}.
On the other hand, the radiative force by diffuse IR photons is always subdominant compared to UV photons in the ionized region due to the smaller dust absorption opacity.
Since the IR photon, however, can propagate into the neutral medium, the resulting radiative force can affect the accretion flow in the non-ionized region.
The 1D simulations by \cite{Toyouchi2019} demonstrate the potential effect of IR radiative force by considering extremely dense environments where the UV radiative feedback no longer works due to the intense ram pressure of accreting neutral gas.
This numerical experiment showed that the IR radiative force significantly regulates the gas accumulation towards the central BH, and the resulting mass accretion rate onto BH is capped with the Eddington value evaluated for the IR radiative force,
\begin{eqnarray}
\meddir = \frac{\leddir}{\eta c^2} \ ,
\label{eq:meddir}
\end{eqnarray}
\begin{equation}
\begin{split}
\leddir &= \frac{4 \pi G \mbh c}{\kappa_{\rm d,IR}} \\
&=  4.4 \times 10^7 \left ( \frac{\mbh}{10^3 \ {\rm M_\odot}} \right ) \left ( \frac{Z}{10^{-2} \ Z_\odot} \right )^{-1} \ L_\odot \ ,
\label{eq:leddir}
\end{split}
\end{equation}
where we adopt $\kappa_{\rm d,IR} = 30 \ (Z/Z_\odot) \ {\rm cm^2 \ g^{-1}}$ supposing the dust sublimation temperature of $T_{\rm d} \sim 1500$~K where the opacity is expected to be highest.
Thus, the IR radiative force could be dominant in neutral regions for $Z \gtrsim 10^{-2}~\zsun$.
In this study, we extend our previous work to non-spherical cases.
In Sections \ref{sec:fiducial} and \ref{sec:wind}, we touch on the effect of the diffuse IR photons on the mass accretion rate onto IMBHs embedded in dense gas disks.


\subsection{Numerical setups and boundary conditions}

In this study, we investigate the accretion flow onto BHs via dusty tori or CNDs that extend from pc to sub-pc scale, hereafter referred as the nuclear region.
Generally, these nuclear disk structures are fed by the mass transportation from the galactic disks extending over kpc scales.
Supposing such an external gas supply, we consider the continuous inward mass flux at the outer boundary,
\begin{eqnarray}
\minf = \fin \medd \ ,
\label{eq:minf}
\end{eqnarray}
where we only consider the cases with $\fin > 1$ to explore the condition for the super-Eddington mass accretion.
The injected gas is assumed to come into through the surface of $|Z| \leq 2 \times 10^4~\rm AU$, 
roughly corresponding to the thickness of cold gas disk with $T \sim 100$ K.
The injection velocity is set to $V_{\rm inf} = 0.5~V_{\rm K,max}$, where the Kepler velocity at the outer boundary is $V_{\rm K,max} = \sqrt{G \mbh / \rmax}$.
We note that the choice of $V_{\rm inf}$ does not affect our simulation results as long as the injection velocity is 
lower than the free-fall value so that the inflowing gas forms a rotationally supported structure before reaching the 
inner-most cell (see also below).

We also assume that the injected gas has a specific angular momentum perpendicular to the equatorial plane, 
the value of which is described as below,
\begin{eqnarray}
\linf = \fk \rmax V_{\rm K,max} = \fk \sqrt{G \mbh \rmax}\ .
\label{eq:linf}
\end{eqnarray}
We set $\fk = 0.5$ throughout this study. 
The choice of $F_{\rm K}$ is motivated by the result of high-resolution, cosmological simulations 
of galaxy formation, where the rotational velocity of a collapsing gas is as large as half of the Keplerian velocity \citep[][]{Abel2002, Yoshida2008}
and indeed the inflow velocity becomes comparable to the rotational velocity \citep{Inayoshi2014a}.
The different choice of $\fk$ has been confirmed to provide no significant impact on the time-averaged mass accretion rates at least in the range of $\fk = 0.3-0.8$.
At the early epoch of the simulation, the angular momentum of injected gas produces a ring-like structure at the centrifugal radius,
\begin{equation}
\begin{split}
\rcent &= \frac{\linf^2}{G \mbh} \\
&= 1.25 \times 10^5~{\rm AU}~\left ( \frac{\fk}{0.5} \right )^2 \left ( \frac{\rmax}{5 \times 10^5~{\rm AU}} \right ) \ .
\label{eq:rcent}
\end{split}
\end{equation}
The ring structure is fed by the gas supply from the outer boundary and eventually fragments due to the self-gravity.
After that, a radially extending disk forms and drives accretion flows toward the central BH.

Note that our simulation assumes constant $\minf$ and $\linf$ throughout the computational time of $\lesssim 10$ Myr.
In reality, the property of injected gas into central pc scales in galaxies can change on the timescale of 1 Myr \citep[e.g.,][]{Hopkins2010}. 
While investigating such more realistic gas inflow history from the galactic disks is essential, in this paper, we aim to acquire the fundamental knowledge for the mass growth of BHs at galactic centers by considering a simple situation.

The computational domain covers the radial range from $\rmin = 5 \times 10^3~{\rm AU}$ to $\rmax = 5 \times 10^5~{\rm AU} \sim 2.5~{\rm pc}$, which resolves the Bondi radius of photo-ionized gas with $T \sim 10^5$ K defined below,
\begin{eqnarray}
\rb &=& \frac{G \mbh}{c^2_{\rm s,\infty}} \\
&=& 1.4 \times 10^4~{\rm AU} \left ( \frac{\mbh}{10^4 \ {\rm M_\odot}} \right ) \left ( \frac{\tinf}{10^5 \ {\rm K}} \right )^{-1} \ ,
\label{eq:rb}
\end{eqnarray}
where we assume the isothermal gas with the polytropic index $\gamma = 1$, the mean molecular weight $\mu = 1.3$, and the sound speed $\cinf = \sqrt{\gamma k_{\rm B} \tinf / (\mu m_{\rm p})} = 8.1 (\tinf/10^4 \ {\rm K})^{1/2} \ {\rm km \ s^{-1}}$. 
We basically assume an equatorial plane symmetry, in which the tangential numerical domains are $0 \leq \theta \leq \pi/2$ and $0 \leq \phi \leq 2 \pi$.
The number of grid cells in each directions are $(N_r, N_\theta, N_\phi) = (128, 36, 72)$ for the basic models introduced in Section \ref{sec:setup}.
We adopt uniform grids in $\phi$ but logarithmic ones in $r$ and $\theta$ to realize higher resolution in the inner and closer region to the equatorial plane, which enables to resolve the disk thickness with at least three grid cells.

\begin{figure*}
\begin{center}
\includegraphics[width=18cm]{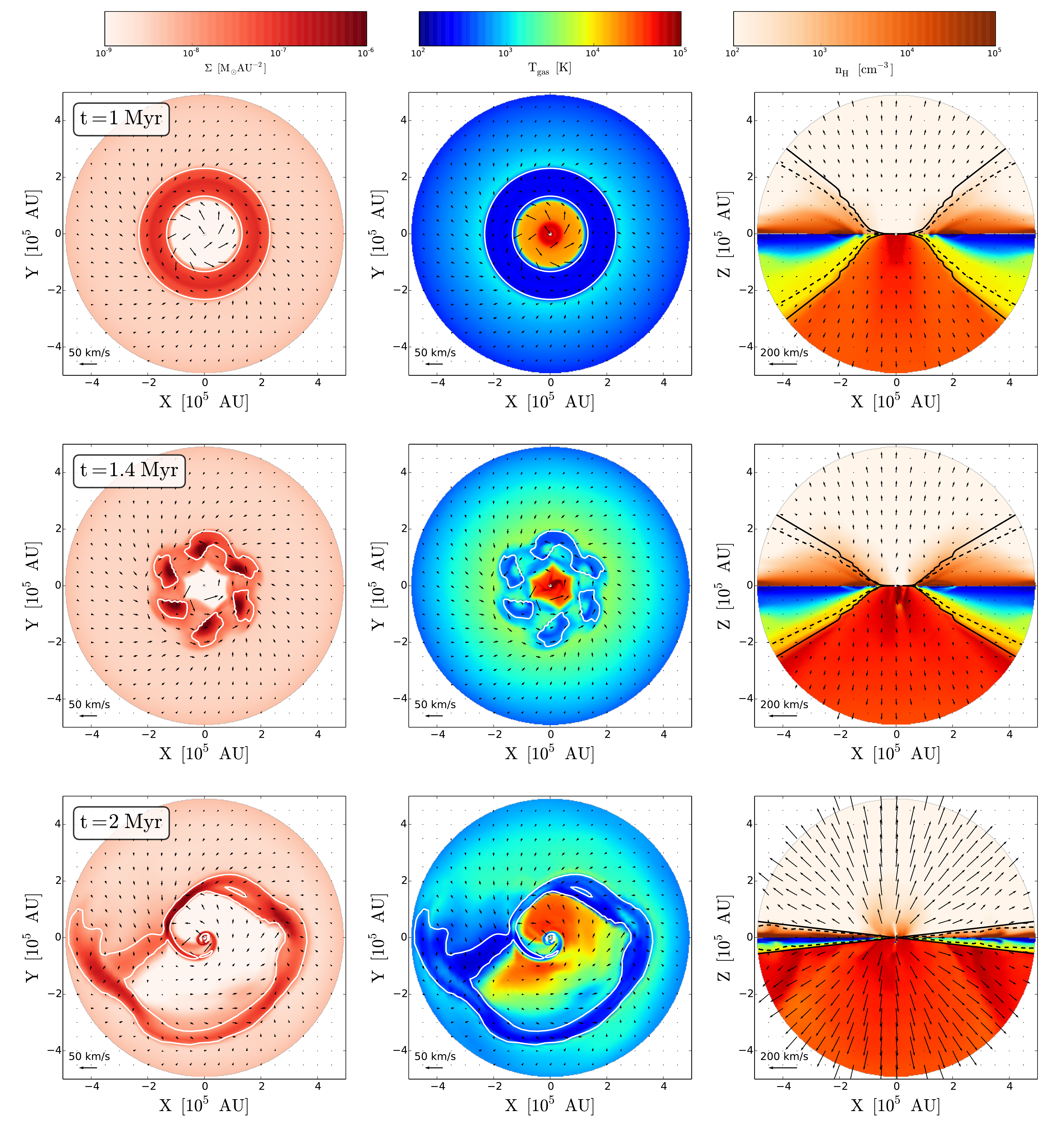}
\end{center}
\caption{Time evolution of the density and thermal structure in the Z-2F2 model. 
The left and middle columns show the face-on distributions of surface mass density and gas temperature averaged over the direction of Z, respectively.
The right column represents the edge-on distributions of gas number density and temperature, which are not averaged but just sliced values in XZ plane.
From the top to the bottom panels, we show the accretion structure at $t =$ 1, 1.4, and 2 Myr. 
The arrows denote the local flow velocities, the reference speed of which is shown at the left bottom corner in each panel. 
The white contours in the left and middle columns indicate the regions where the Toomre-Q value is $Q \equiv c_{\rm s} \Omega / (\pi G \Sigma) = 1$.
In the right column, the sold and dashed curves denote the layers where the neutral fraction o gas is 0.01 and 0.99.
At the beginning of the simulation, injected gas from the outer boundary forms a ring structure
and fragments into clumps due to its self-gravity of gas at $t \sim$ 1.4 Myr.
As a result of ring fragmentation, mass accretion onto the central BH efficiently proceeds and more intense radiation output from the accreting BH leads to a significant amount to mass loss from the disk owing to photoevaporation.}
\label{fig:str_Z2F2}
\end{figure*}


\subsection{Models} \label{sec:setup}

With RHD simulations, we investigate the gas accretion onto the IMBH with $\mbh = 10^4~\msun$, supposing a growing seed BH in the early universe. 
Table \ref{table:model} summarizes the models presented in this paper. 
The top seven models, hereafter called the basic ones, aim to explore the dependence of the mass accretion efficiency of the IMBH on $Z$ and $\fin$. 
The first three and second two characters in the name of each model give the logarithmic value of $Z$ and $\fin$, respectively: our fiducial model with $Z = 10^{-2}~\zsun$ and $\fin = 10^2$ is tagged as Z-2F2.

We study the metallicity range of $Z=10^{-3}-10^{-1}~\zsun$, supposing the interstellar medium in the early galaxies where the chemical enrichment has not proceeded well. 
On the other hand, we assume $\fin$ = 10-1000, roughly corresponding to the injection rate from the outer boundary of $\minf \sim 10^{-3}-10^{-1}~\msun \rm yr^{-1}$. 
Realizing this gas supply rate is possible because it is much smaller than that observed in the hydrodynamical simulation of the Milky Way like galaxy-galaxy merger by \cite{Hopkins2010}. 
In Section \ref{sec:possible}, we argue when and where the situations supposed here appear in the context of galaxy evolution.

In addition to the seven basic models, we present some experimental calculations based on our fiducial model, Z-2F2. 
Z-2F2nuv and Z-2F2nir take into account no radiative force from direct UV and diffuse IR photons, respectively.
For checking effects of numerical resolution and grid configuration, Z-2F2hr adopts the twice number of grids in $\theta$, Z-2F2ne relaxes the assumption of the equatorial symmetry, and Z-2F2hr+ne is a combination of these two models.
Moreover, we perform Z-2F3hr+ne corresponding to Z-2F2hr+ne but based on Z-2F3 model.
In the next section, we show the results of these seven basic and six test models.


\section{RESULTS} \label{sec:result}

\subsection{Overview of numerical results} \label{sec:mdot}

\subsubsection{Fiducial (Z-2F2) case} \label{sec:fiducial}

Figure \ref{fig:str_Z2F2} shows the density and temperature distribution of the accretion flow obtained in our fiducial model (Z-2F2). 
At $t =$ 1 Myr, the injected gas from the outer boundary settles in a ring structure around its centrifugal radius. 
The gas ring has grown with time and eventually fragments due to self-gravity of the gas at $t \sim 1.4$ Myr, 
which forms a radially extending disk. 
At $t =$ 2 Myr, a non-axisymmetric spiral-arm structure forms and interacts with the surrounding gas, 
leading to efficient angular momentum transport and mass accretion onto the BH.
Since the rapidly accreting BH releases intense radiation preferentially toward the polar regions (see Eq. \ref{eq:aniso_flux}),
the gas above the disk height is ionized and evacuated.
However, mass inflow is still allowed through the equatorial region where the gas density is high enough to 
cool via metal line emission and maintain the marginally-unstable disk structure.

\begin{figure}
\begin{center}
\includegraphics[width=8cm]{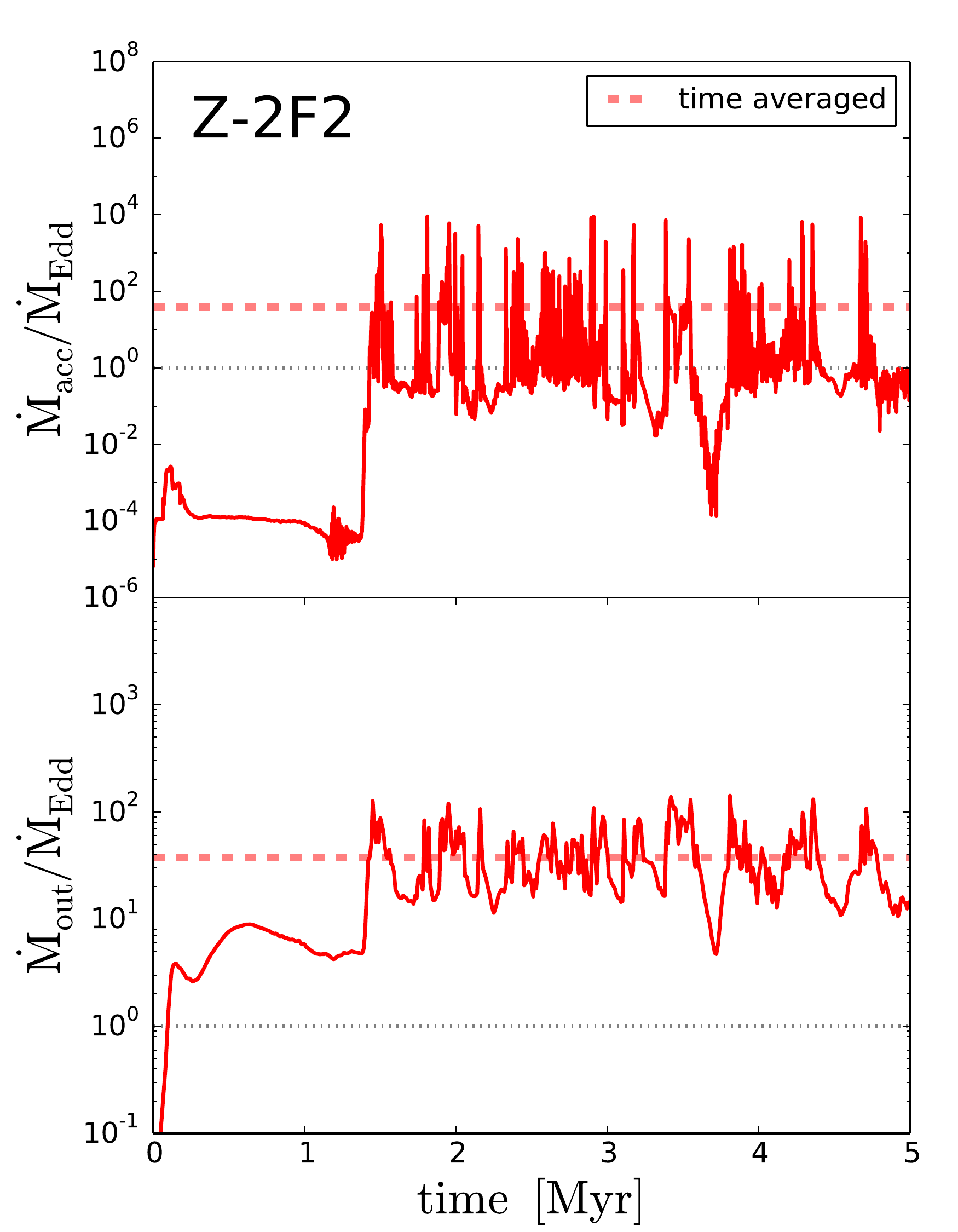}
\end{center}
\caption{Time evolution of mass accretion (top) and outflow rates (bottom) obtained in our fiducial model, Z-2F2.
The vertical axis shows the values normalized by the Eddington accretion rate (dotted lines). 
Dashed lines represent the time-averaged mass accretion and outflow rates.
After the ring fragmentation around $t \sim$ 1.5 Myr, 
both the mass accretion and outflow rates rapidly rise and begin to oscillate with time.
}
\label{fig:mdot_Z2F2}
\end{figure}

The time variability of the mass accretion rate in our fiducial case is shown in the top panel of Figure \ref{fig:mdot_Z2F2}. 
As expected from Figure \ref{fig:str_Z2F2}, the mass accretion rate drastically rises due to fragmentation of the ring 
at $t \sim 1.4$ Myr, and begins to oscillate with peak values at $\sim 10^4~\medd$.
The intermittent accretion behavior is caused by the state transition of the disk between gravitationally stable and unstable phases.
In the quiescent phase, where the disk is stable and angular momentum transport is inefficient, the injected mater from the outer
boundary is accumulated.
Once the disk becomes massive enough to be gravitationally unstable, a large amount of the gas can fall into the central BH 
due to efficient angular momentum transport by the non-axisymmetric structures.
Since the burst-like accretion reduces the surface density of the disk, the gravitational instability is self-regulated and thus 
the disk results in a quiescent phase.
In fact, the typical interval of accretion bursts seems consistent with the orbital timescale of $\sim 0.1$ Myr
at the centrifugal radius of $\sim 10^5$ AU
\footnote{
We have conducted several different simulations setting larger centrifugal radii of the injected gas, 
where the typical interval timescale between bursts becomes longer but the time-averaged properties 
of the accreting flow are hardly affected.  
}.
Even with such short variabilities, however, the time-averaged accretion rate is $\mdott \sim 40 \medd$, 
as shown in Figure \ref{fig:mdot_Z2F2} (dashed line).
This implies that the BH would increase its mass by a factor of $\sim 5$ within $\sim$ 5 Myr via
super-Eddington accretion (although the BH mass is fixed through this simulation).
Note that those rapidly growing phases would not last long because the disk tends to be gravitationally stable
as the BH mass is higher than the disk mass and the shear velocity increases in the disk.

\begin{figure*}
\begin{center}
\includegraphics[width=16cm]{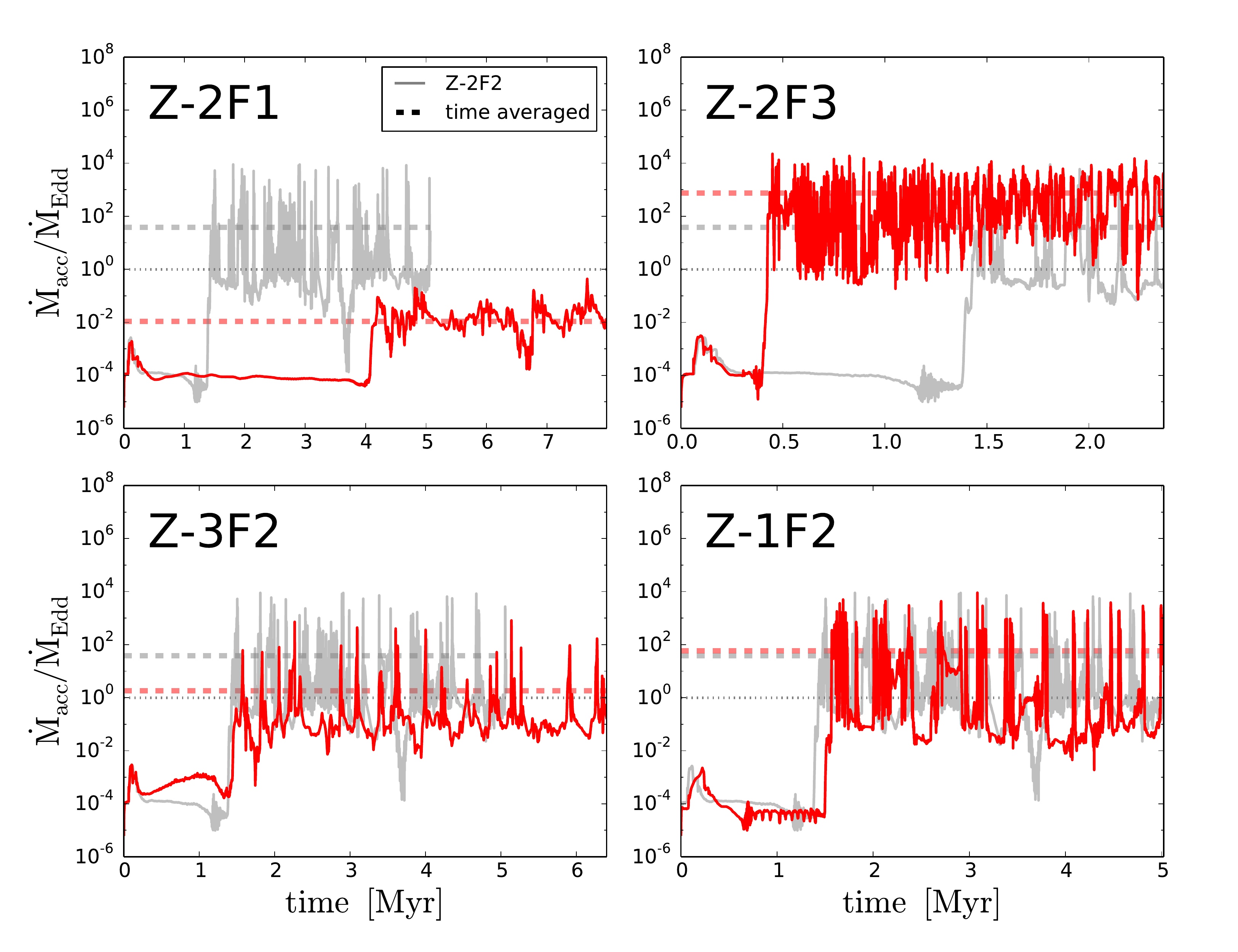}
\end{center}
\caption{Time evolution of mass accretion rates obtained in the four basic models, Z-2F1, Z-2F3, Z-3F2, and Z-1F2.
For comparison, the result of the fiducial Z-2F2 model is overlaid in each panel (grey curve).
}
\label{fig:mdot}
\end{figure*}

A similar episodic behavior of the mass outflow rate at the outer boundary is seen in the bottom panel of Figure \ref{fig:mdot_Z2F2}.
The outflow rate tightly correlates with the mass accretion rate over time, implying that radiative feedback 
mainly drives wind mass loss from the disk surface.
The time-averaged outflow rate is as high as $\moutt \sim 40~\medd $, which is comparable to the BH accretion rate of $\mdott$.
Since the outflowing matter is launched from larger radii, where the dynamical timescale is longer, 
the fluctuations of $\mout$ are quite modest compared to those of $\mdot$.
We note that $\sim 20\%$ of the gas injected from the outer boundary stays in the disk and makes it gravitationally unstable;
namely the disk mass is $ \sim 0.2~ \minf \times 5~{\rm Myr} \sim 2 \times 10^{4}~\msun$.
We note that the disk mass reaches an almost constant value by the end of the simulation
and the accretion system has been in a quasi-steady state where $\mdot \simeq \minf - \mout$ is satisfied.

Here, we investigate the effects of radiative force on the accretion dynamics with Z-2F2nuv and Z-2F2nir models, where the radiative force due to direct UV and diffuse IR photons are ignored, respectively.
As shown in Table \ref{table:model}, Z-2F2nuv does not significantly differ in $\mdot$ and $\mout$ from the fiducial case.
This result implies that the strong outflow is caused by photoevaporation of accreting gas rather than the radiation force through electron scattering and dust absorption of UV photons.
In fact, the radiation force does not exceed the gravitational force from the BH near the equatorial plane due to anisotropic radiation. 
Therefore, the direct UV irradiation is not intense enough to repel the accretion flows.
Thus, anisotropic radiation reduces the negative impacts of the radiation force on the disk dynamics and assists super-Eddington mass accretion onto the BH.
On the other hand, in a dusty accretion disk, IR radiation generally could affect the disk dynamics because IR photons penetrate even near the equatorial plane in a diffusive way via absorption and re-emission by dust grain.
However, Z-2F2nir model indicates that the IR radiative force also provides no significant impacts on the accretion dynamics.
This is because the optical depth within the disk is less than unity, and therefore diffusive IR photons cannot be trapped within the accretion disk.
Note that such an optically thin limit can break for cases with higher values of $\fin$ and $Z$.
In Section \ref{sec:wind}, we further discuss the effects of IR radiative force on mass transfer along accretion disks.

\begin{figure}
\begin{center}
\includegraphics[width=8cm]{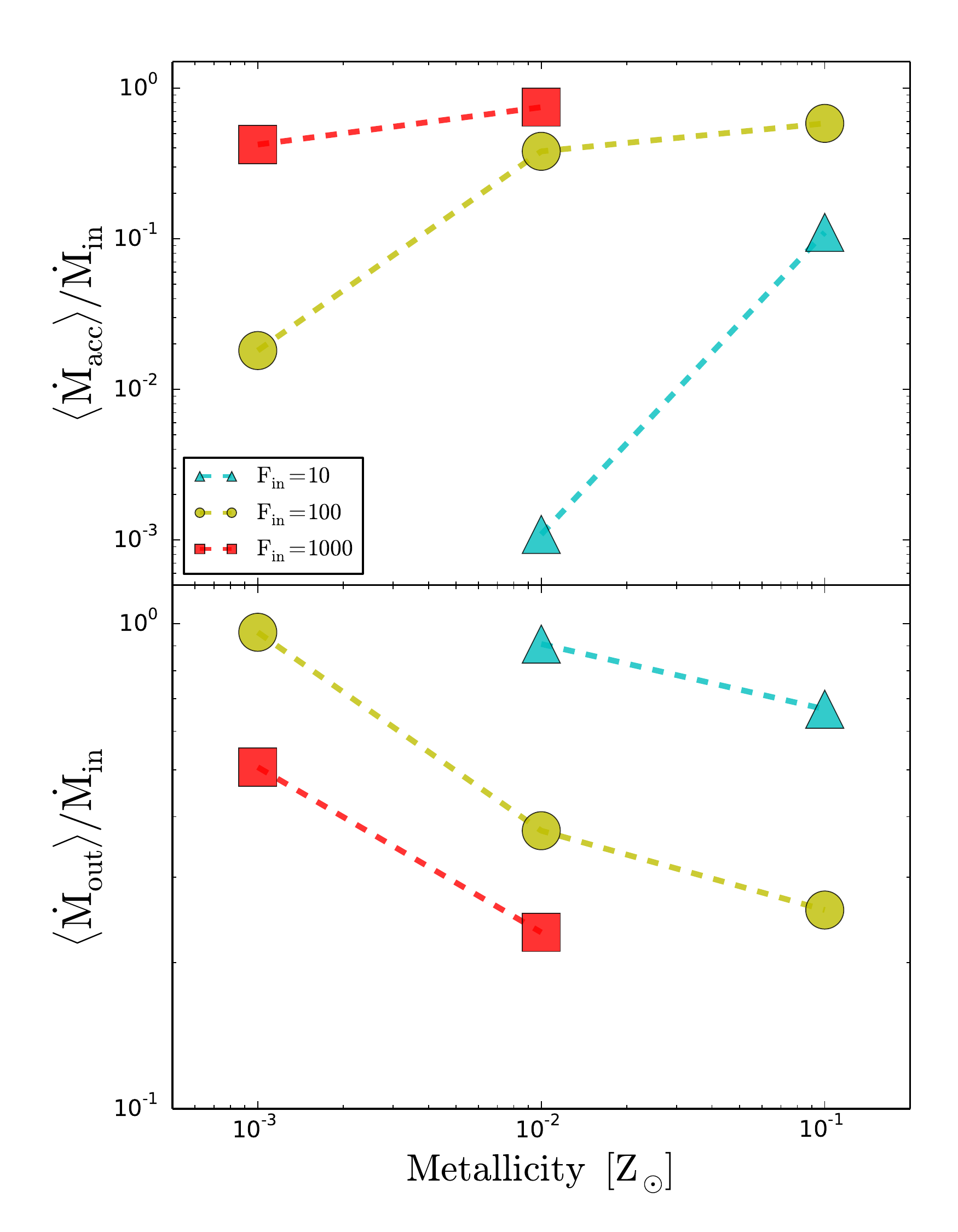}
\end{center}
\caption{Variations of the mass accretion and outflow efficiencies 
for different metallicity $Z$ and mass injection parameter $\fin$.
The top and bottom panels indicate the time-averaged mass accretion and outflow rates normalized by 
the injection rates from the outer boundary as a function of gas metallicity.
Cyan triangles, yellow circles, and red squares correspond to the cases with $\fin =$ 10, 100, and 1000.
The efficiency of mass accretion (outflow) increases (decreases) for higher $Z$ and $\fin$.
}
\label{fig:mdot_summary}
\end{figure}

\begin{figure*}
\begin{center}
\includegraphics[width=15cm]{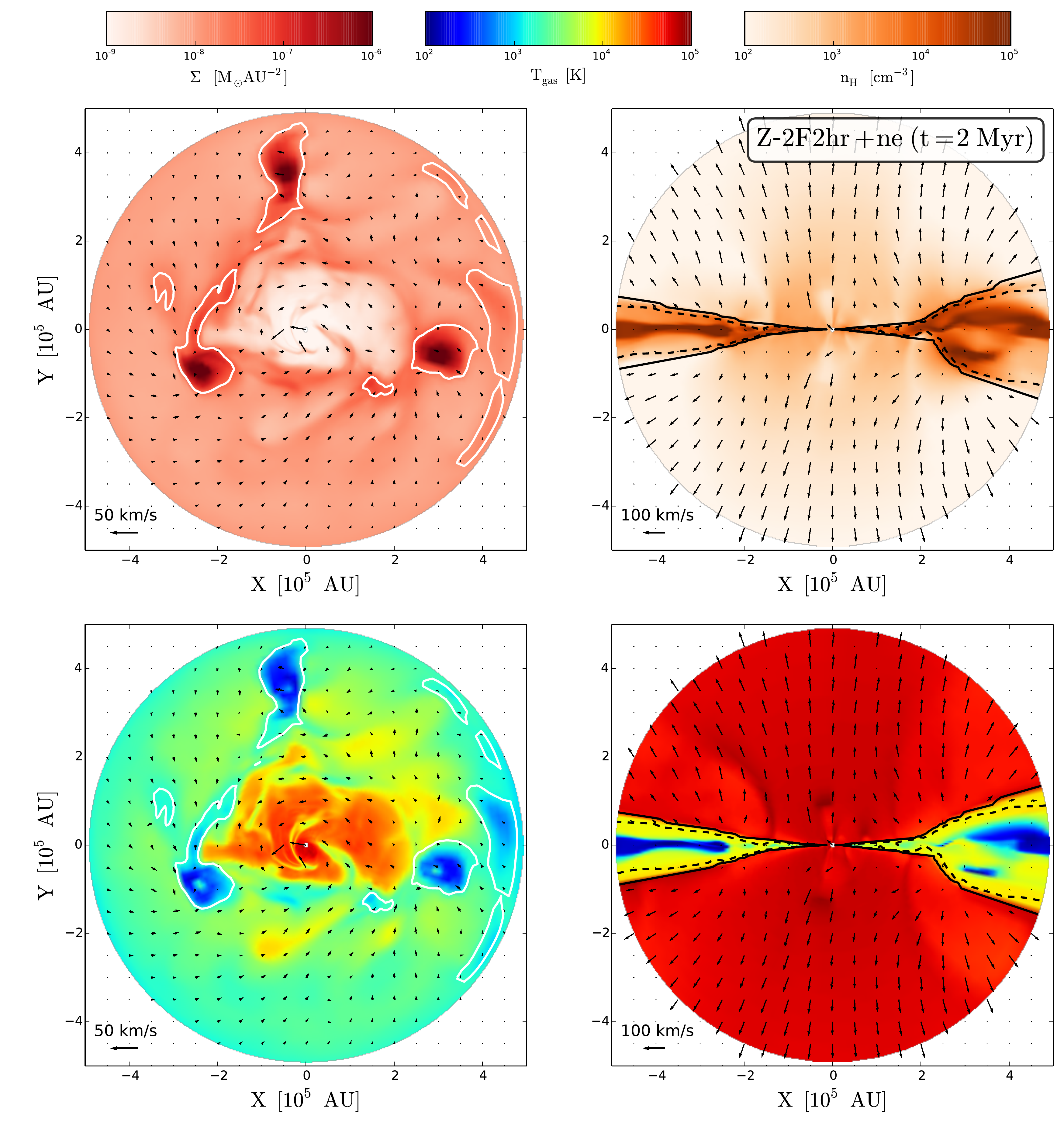}
\end{center}
\caption{Same as the bottom panels in Figure \ref{fig:str_Z2F2}, but for the Z-2F2hr+ne model,
where the resolution in the $\phi $ direction is twice higher than that in the Z-2F2 case and 
the equatorial-plane symmetry is relaxed.
Since the disk structure becomes more clumpy and clumps move in the vertical direction,
the effective thickness of the disk increases.
}
\label{fig:str_Z2F2hr}
\end{figure*}

\begin{figure}
\begin{center}
\includegraphics[width=8cm]{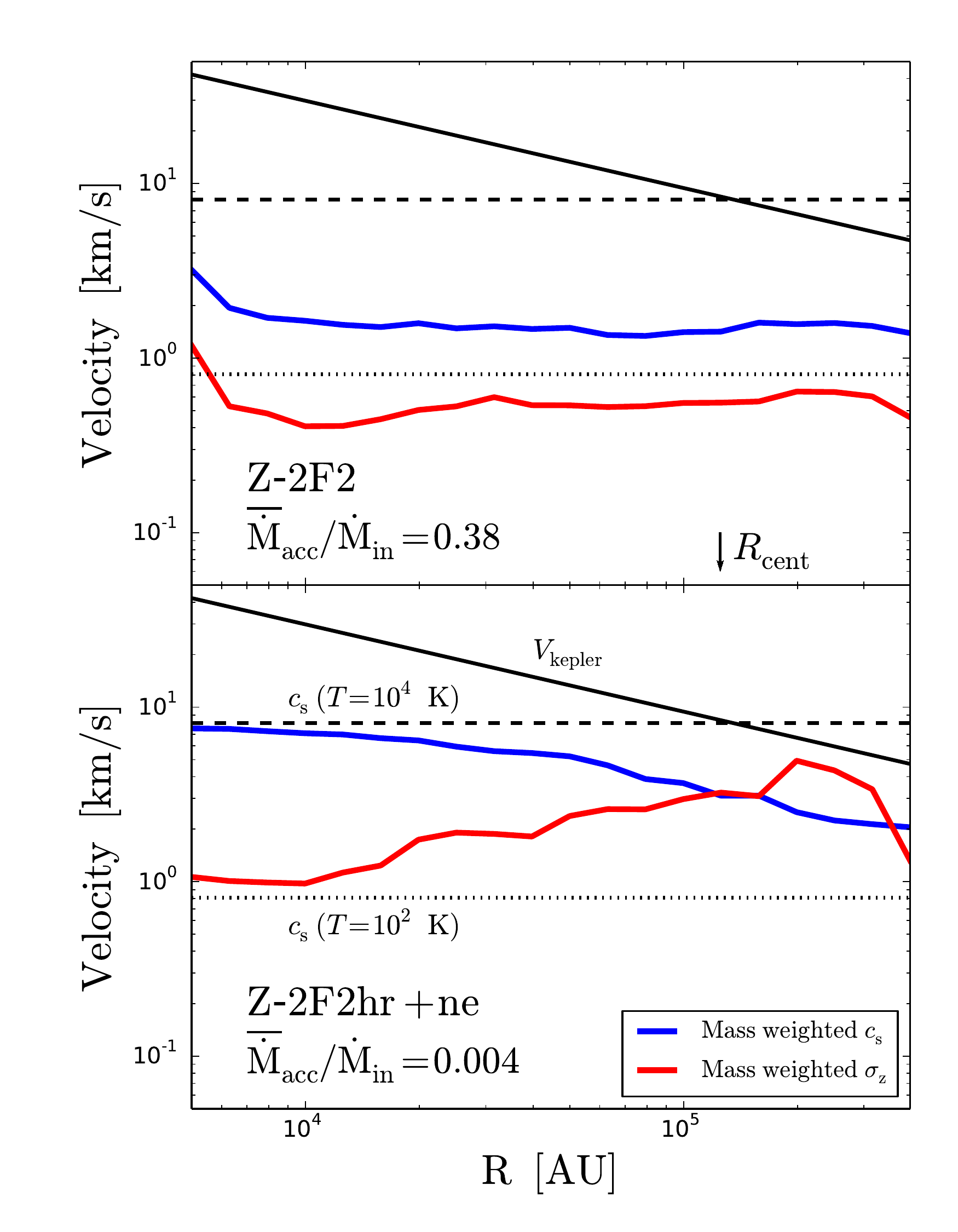}
\end{center}
\caption{Radial profiles of velocity components in the Z-2F2 (top) and Z-2F2hr+ne (bottom) models, respectively.
The blue and red curves represent the mass-weighted sound speed and velocity dispersion in the vertical direction.
Those quantities are evaluated for neutral gas with $T \sim 10^4$ K and are averaged over time after the epoch of ring fragmentation.
For comparison, the Kepler velocity (solid), and sound speed for $T = 10^4$ K (dashed) and $10^2$ K (dotted) are presented.
The location of the centrifugal radius of injected gas is indicated with the arrow.
In the Z-2F2hr+ne model, both the kinetic velocity and sound speed are substantially higher. 
}
\label{fig:E_r}
\end{figure}

\begin{figure}
\begin{center}
\includegraphics[width=8cm]{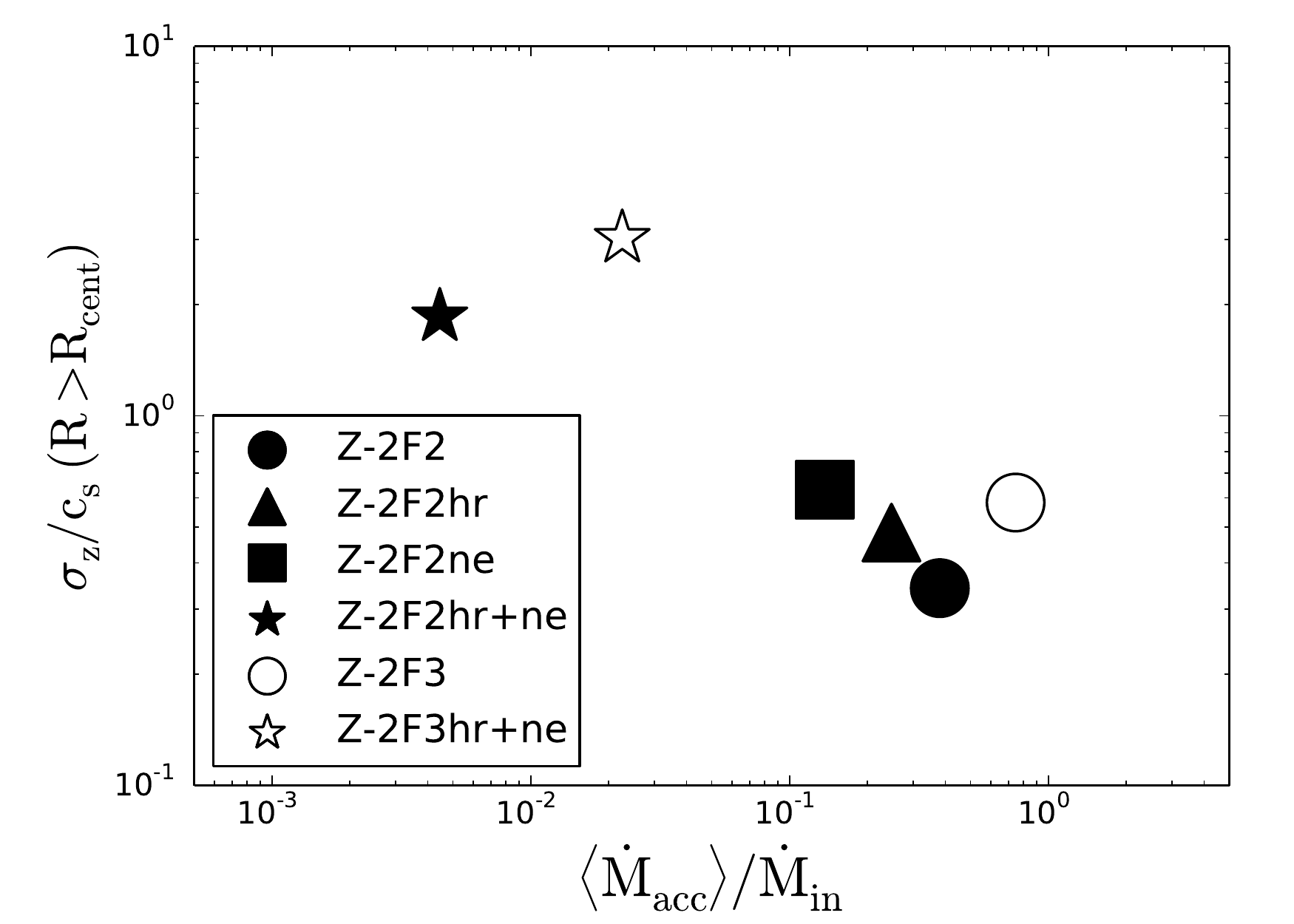}
\end{center}
\caption{The correlation between the mass accretion efficiency and the ratio of 
$\sigma_{\rm z}/c_{\rm s}$ (volume averaged at $R > \rcent$).
The plotted values are the time-averaged value after the epoch of ring fragmentation in each model.
Different symbols show the results of different models shown in the figure.
In the Z-2F2hr+ne and Z-2F3hr+ne models (star symbols), 
mass accretion is reduced in dynamically hotter disks.
}
\label{fig:hz_E_NC}
\end{figure}

\subsubsection{Dependence on metallicity and mass injection rate} \label{sec:dependence}

We here investigate the effect of varying the metallicity $Z$ and the mass injection rate $\fin$ on the properties of accretion flows. 
Figure \ref{fig:mdot} presents the mass accretion rates for the four models with various values of $Z$ and $\fin$
(Z-2F1, Z-2F3, Z-3F1, and Z-1F1).
For comparison, we overlay the result of the fiducial model (gray curve) in each panel
(note that the simulation terminates at different times).
In Figure \ref{fig:mdot_summary}, we summarizes the mass accretion and outflow rate normalized by the mass injected rate, 
respectively (see also Table \ref{table:model}).

First, we compare the two models of Z-2F1 ($\fin =10$) and Z-2F3 ($\fin =10^3$) with 
the same metallicity as in the fiducial case.
With the higher injection rate, ring fragmentation occurs earlier and the BH is fed at 
a higher accretion rate of $\mdott \sim 750~\medd$, which is $\sim 20$ times higher than that 
in the fiducial case.
This indicates that the ratio of the mass accretion rate to the injection rate from the outer boundary 
increases to $\sim 75 \%$.
With the lower injection rate, the transition of mass accretion occurs later because it takes a longer time
for the disk to become unstable.
Unlike the other case, the mass accretion is much lower than the value expected from the ratio of $\fin$;
namely $\mdott \simeq 10^{-2}~\medd$, which is only $\sim 0.1 \%$ of the injected gas.
This is because in this case, $\sim 90\%$ of the injected mass is ejected from the disk as winds.

Next, we discuss the other two models of Z-1F2 ($Z=0.1~\zsun$) and Z-3F2 ($Z=10^{-3}~\zsun $) with 
the same mass injection rate as in the fiducial case.
In the higher metallicity case, the overall behavior of mass accretion rate is similar to that in the fiducial case,
except the higher value of $\mdott \sim 60~\medd$.
In contrast, with the lower metallicity, the accretion rate is suppressed and the time-averaged rate is limited at $\mdott \sim 2~\medd$.
This result is opposite to the previous 1D RHD simulations of BH accretion where 
the radiation force onto dust grain prevents mass accretion onto BHs as the metallicity increases
\citep[e.g.,][]{Yajima2017a, Toyouchi2019}.
Therefore, the anisotropy of radiation field and geometrical effect qualitatively change the accretion dynamics
at super-Eddington rates.

As summarized in Figure \ref{fig:mdot_summary} (including two more cases of Z-3F3 and Z-1F1),
higher values of $Z$ and $\fin$ lead to higher mass accretion rates onto BHs,
suppressing mass outflows from the disk.

\begin{figure*}
\begin{center}
\includegraphics[width=13cm]{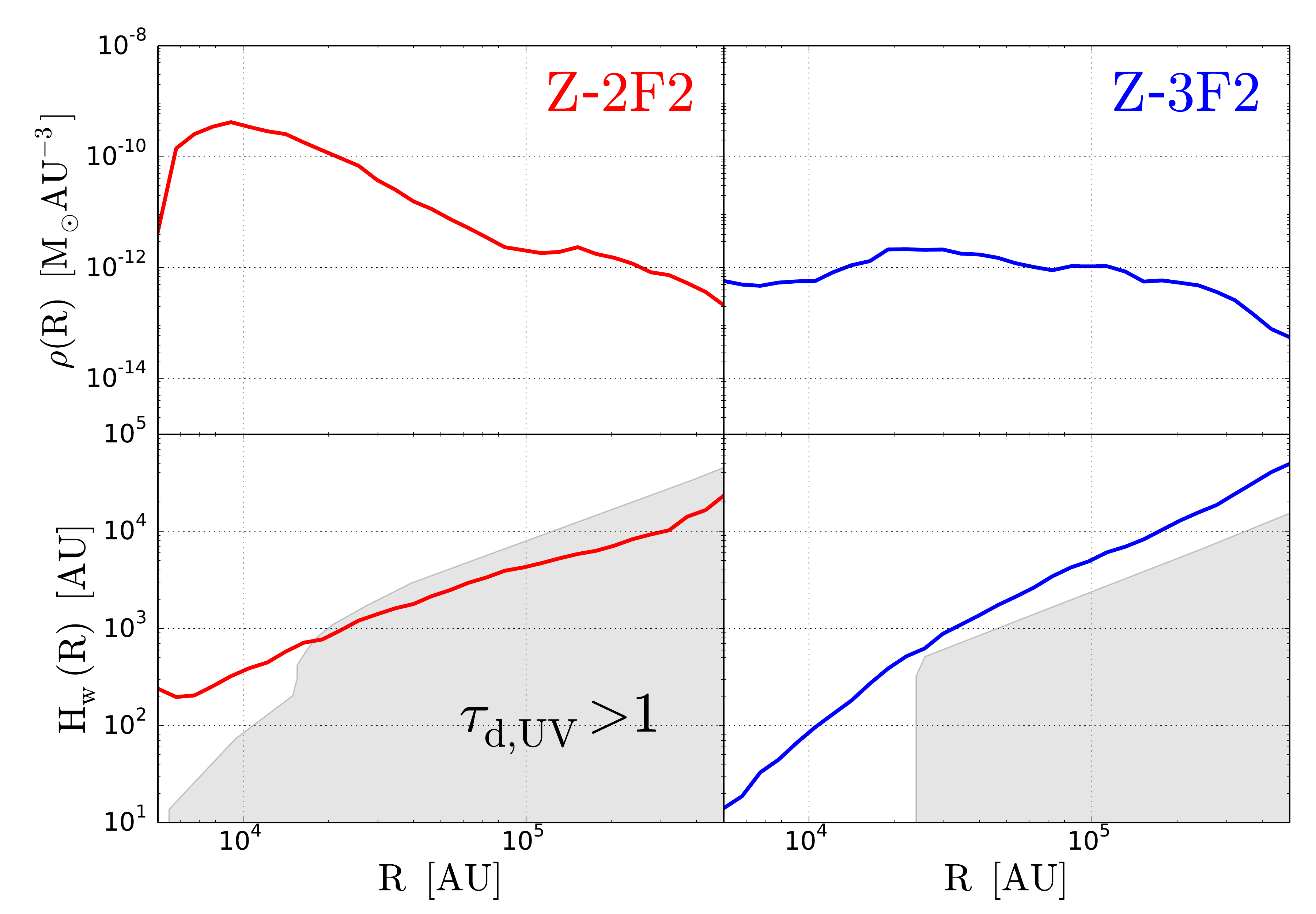}
\end{center}
\caption{Geometrical differences of the azimuthally and time-averaged ($0\leq \phi \leq 2\pi$ and $t \geq 1.5$ Myr) 
disk properties for the Z-2F2 (left) and Z-3F2 (right) models.
The top and bottom panels show the radial profiles of the gas density vertically averaged within the neutral region 
and the disk scale height derived with Eq. (\ref{eq:hz_wm}), respectively.
In the bottom panel, grey shaded area indicates the optically thick regime for dust absorption of ionizing photons.
For the higher metallicity model a dense gaseous disk forms with minor effects of radiative feedback, 
whereas for the lower metallicity one the accretion disk is so optically thin to suffer from significant mass loss via photoevaporation.
}
\label{fig:den_hz}
\end{figure*}

\begin{figure*}
\begin{center}
\includegraphics[width=16cm]{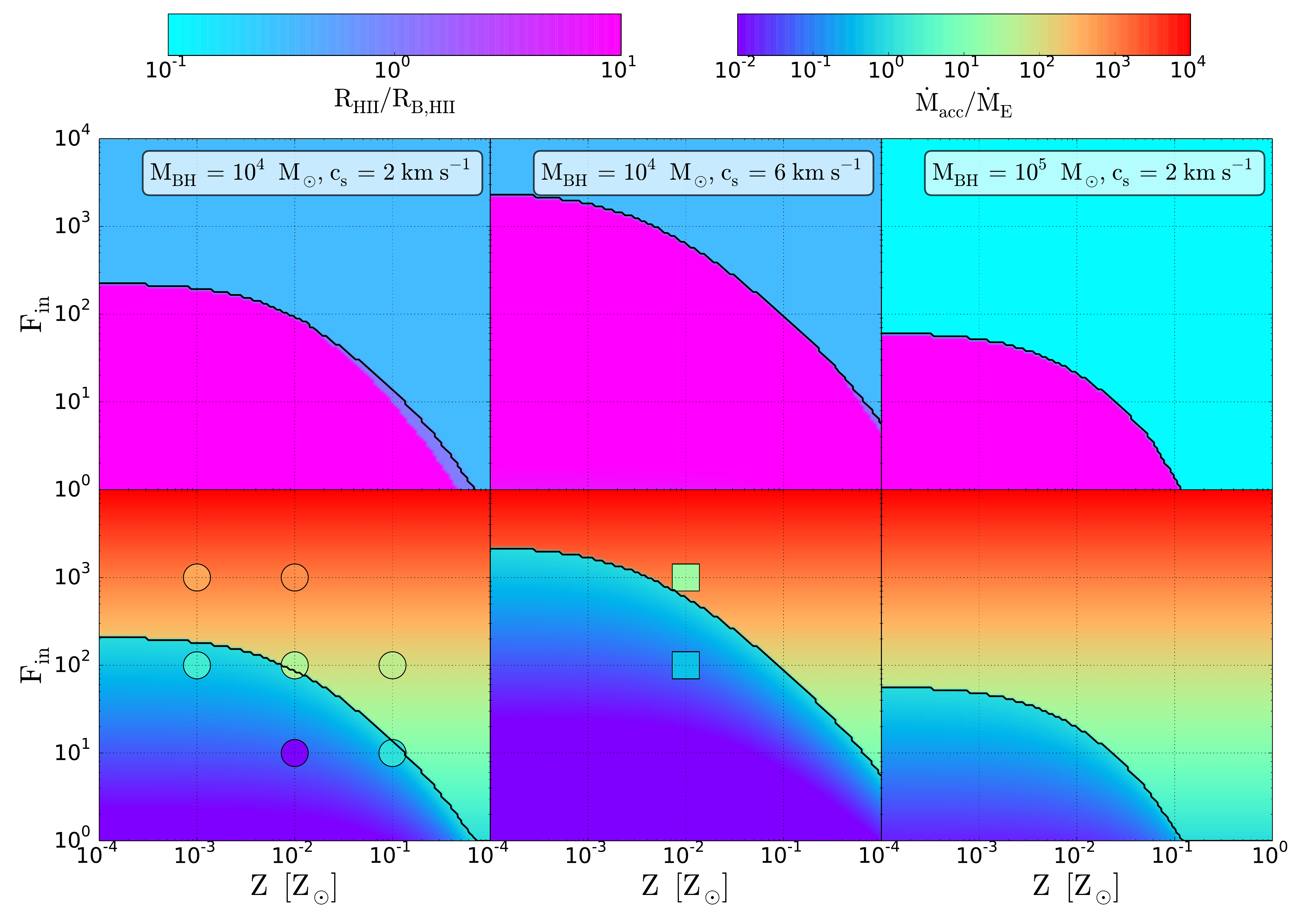}
\end{center}
\caption{The equilibrium values of $\rioni$ and $\mdot$ derived with the 1D accretion disk model are indicated with the top and bottom colormaps, respectively.
Each columns shows the case with ($\mbh,~c_{\rm s}$) = ($10^4~\msun,~2~\rm km~s^{-1}$), ($10^4~\msun,~6~\rm km~s^{-1}$), and ($10^5~\msun,~2~\rm km~s^{-1}$), respectively.
The black solid curves in the top and bottom panels indicate the boundary between the efficient and inefficient accretion modes;
namely, $\rioni = R_{\rm B,HII}$ and $\mdot = \medd$, respectively.
For comparison, colored circles overplotted in the left column present the mass accretion rates obtained in the seven basic models, and squares in the middle column correspond to Z-2F2hr+ne and Z-2F3hr+ne.
Our 1D model predicts that higher values of $\fin$ and $Z$ lead to higher accretion rates, which is generally consistent with the results of our RHD simulations.}
\label{fig:rst}
\end{figure*}

\subsection{Numerical Convergence} \label{sec:nc}

In this section, we check the numerical convergence of our simulation results,
performing two higher-resolution runs with $(N_\theta, N_\phi) = (72, 144)$ and relaxing the equatorial-symmetry assumption
(Z-2F2hr+ne and Z-2F3hr+ne).

Figure \ref{fig:str_Z2F2hr} presents the density and temperature structure of the Z-2F2hr+ne model (i.e., $Z=10^{-2}~\zsun$ and $\fin=100$).
The face-on views (left panels) show an accretion disk with more clumpy structures compared to the fiducial case,
because non-axisymmetric spiral arms further fragment into those smaller clumps in the higher-resolution model. 
In contrast, the inter-clump region is too rarefied to cool down to $T \lesssim 10^4$ K via CII and OI fine-structure lines
against dynamical heating in the disk.

The edge-on views (right panels) present that the accretion disk 
becomes substantially thicker than that in the fiducial case,
because the equatorial symmetry is relaxed and the massive clumps dynamically heat the disk.
To describe this quantitatively, we show the radial profiles of mass weighted sound speed $c_{\rm s}$ and vertical velocity 
dispersion $\sigma_z$ for the Z-2F2 and Z-2F2hr+ne models in Figure \ref{fig:E_r}.
In the high-resolution case, the gas temperature is higher everywhere and the kinetic velocity in the $z$-direction 
exceeds the sound speed outside the centrifugal radius where dense clumps form, although the disk in the fiducial case is
dynamically and thermally colder.
We note that the clump properties are characterized with the mass weighted quantities because 
the dense clumps orbiting at larger radii dominates the mass.
Since dense clumps move sub-sonically in warm gas with $T \gtrsim 10^4$ K, the vertical motion of those clumps 
does not form shocks between the clumps and interclump media.
In addition, the vertical velocity dispersion in the Z-2F2hr+ne model asymptotically approaches the Keplerian velocity in the outer regions.
This suggests that a fraction of the rotational energy injected from the outer boundary is re-distributed to the vertical kinetic motion
and supports the clumpy disk in the vertical direction.

Gas kinetic motion in the disk affects mass accretion flows to central BH. 
Figure \ref{fig:hz_E_NC} shows a negative correlation between the mass accretion efficiency and 
the ratio of $\sigma_z/c_{\rm s}$ measured outside the centrifugal radius.
Namely, the values of $\mdott/\minf$ in the Z-2F2hr+ne and Z-2F3hr+ne models are reduced by a factor of $\sim 50-100$
from those in their counterpart (lower-resolution and equatorial symmetry) models, as 
the disk becomes dynamically hotter.
We also present the results of Z-2F2hr and Z-F2ne models, in which either of the higher resolution or 
non-equatorial symmetry is considered.
For these two models, the reduction of the mass accretion efficiency from the fiducial case is quite modest.
This fact implies that the combination of clump formation and non-zero vertical motions of gas is 
essential to affect the disk accretion dynamics.
Thus, our simulation results still depend on the numerical resolution and grid configuration.
However, even in the higher resolution cases, the qualitative dependence of mass accretion rates on the mass injection rate holds, and rapid accretion exceeding the super-Eddington rate is feasible when $\fin > 1000$.
In next section, we provide more detailed arguments about the physical origin of the dependence on $Z$, $\fin$, and the kinetic velocity.


\section{CONDITION FOR RAPID MASS ACCRETION} \label{sec:condition}

\subsection{Implication from Simulation Results} \label{sec:imply}

In this section, we provide a simple analytic argument for the conditions required for rapid mass accretion. 
As we described, the mass accretion efficiency increases with the mass injection rate and metallicity (in \S\ref{sec:dependence}),
and also depends on the thickness (i.e., the vertical velocity dispersion) of a gravitationally unstable disk (in \S\ref{sec:nc}).
For this purpose, we study the properties of the accretion disk structure and attenuated radiation field in more details,
specifically focusing on the differences in the two cases of Z-2F2 and Z-3F2
as representative models that show different accretion efficiencies due to metallicity effects.
For convenience, we introduce the cylindrical distance of $R = r \sin \theta$ to describe the disk properties
in the following discussion.

In the top panels of Figure \ref{fig:den_hz}, we present the radial profiles of the time-averaged gas density 
of neutral gas within the accretion disk for the two models. 
With the higher metallicity (Z-2F2 model), the gas density continuously increases inward and reaches 
$\sim 10^{-9}~\msun~\rm AU^{-3}$ near the center, although it declines near the inner boundary.
Overall, the density profile follows $\rho(R) \propto R^{-2}$, 
which indicates the inflow velocity is characterized by the free-fall velocity 
($v_{\rm R}\sim v_{\rm ff} \equiv \sqrt{2GM_{\rm BH}/R}$ 
and thus $\dot{M}(R)\simeq R \Sigma v_{\rm R} \simeq (\rho c_{\rm s} / \Omega) v_{\rm R} \propto \rho R^2$) 
due to efficient angular momentum 
transport in the marginally unstable disk.
In contrast, with the lower-metallicity (Z-3F2model), such a dense gaseous disk does not form, 
but the density is saturated almost at $\sim 10^{-12}~\msun~\rm AU^{-3}$ within the centrifugal radius.
Hence, the density is expressed with the mass flux through the disk $\dot{M}(R)$ as
\begin{eqnarray}
\rho(R) 
\simeq  \frac{\dot{M}(R)}{4 \pi c_{\rm s} R^2}
\simeq  \frac{\fin \medd - \dot{M}_{\rm out}(>R)}{4 \pi c_{\rm s} R^2} \propto R^{-\alpha}\ ,
\label{eq:rhocent}
\end{eqnarray}
where $\minf = \fin \medd$ is the mass injection rate, and $\dot{M}_{\rm out}(>R)$ is the mass outflow rate 
integrated over $>R$.
When the mass-loss rate is negligible, one obtain $\alpha = 2$.
Therefore, the density slope becomes shallower (i.e., $\alpha<2$) when the disk mass is removed due to 
radiative feedback associated with BH accretion.
Thus, the higher metallicity model loses only small amount of gas from the inner limited region, whereas the lower metallicity one suffers from significant outflows from the whole disk (see also Figure \ref{fig:mdot_summary}).

In the bottom panels of Figure \ref{fig:den_hz}, we present the radial profiles of the time-averaged disk scale height for the Z-2F2 and Z-3F2 models,
where the height is defined for warm neutral gas with $T \sim 10^3$-$10^4$ K as
\begin{eqnarray}
H_{\rm w} = \sqrt{ \frac{\sum_{i} z_i ^2 \rho_{{\rm w},i} \Delta V_i}{\sum_{i} \rho_{{\rm w},i} \Delta V_i} } \ ,
\label{eq:hz_wm}
\end{eqnarray}
where $z_i$ is the height from the equatorial plane, 
$\rho_{{\rm w},i}$ is the density of warm neutral gas, 
and $\Delta V_i$ is the volume element at the $i$-th grid cell.
The summation in Eq. (\ref{eq:hz_wm}) is taken only over the cells where the temperature is in the range of $10^3~{\rm K} \leq T \leq10^4~{\rm K}$.
With the lower metallicity, the disk becomes substantially thinner in the inner region, where 
the disk mass undergoes photoevaporation, resulting in a smaller amount of neutral gas 
left in the equatorial region.
To explicitly show radiation attenuation into the inner disk, 
we present the region where the optical depth to ionizing photons against 
dust absorption becomes above unity in Figure \ref{fig:den_hz} (shaded regions).
Here, the optical depth is calculated as,
\begin{eqnarray}
\tau_{\rm d, UV}(r, \theta) = \int^r_{\rmin} \kappa_{\rm d,UV}\cdot  \rho(r', \theta) ~{\rm d}r' \ .
\label{eq:tau_uv}
\end{eqnarray}
For the Z-2F2 model, since the disk thickness agrees to the boundary of $\tau_{\rm d, UV} = 1$ at all the radii, 
most of the disk region is shielded to ionizing photons produced from the accreting BH.
In contrast, for the Z-3F2 model, the disk height is well above the optically-thick region and 
the inner region ($R < 2 \times 10^4~\rm AU$) is heated by unattenuated radiation,
driving a significant amount of mass loss.
In fact, this trend of the relative position between the dust photosphere and disk height holds 
for all other cases: rapid accretion models (Z-1F2, Z-3F3, Z-2F3, Z-1F1) and inefficient accretion models
(Z-2F1, Z-2F2hr+ne, Z-2F3hr+ne).
Therefore, formation of a dense disk shielded by dust grain is required to achieve rapid mass accretion onto BHs.

\subsection{Consideration with One-Dimensional Disk Model} \label{sec:consider}

Next, we present a one-dimensional semi-analytical model to quantify the penetration of ionizing photons into the disk 
and estimate the equilibrium mass accretion rate resulting from the photoevaporation effect.
In the 1D model, the density profile is described with Eq. (\ref{eq:rhocent}), 
and for a given mass accretion rate $\mdot$ and size of an ionized region $\rioni$, 
the density slope is assumed to be
\begin{eqnarray}
\alpha(R) =  
\begin{cases}
2 - \frac{{\rm log} \left ( \mdot / \minf \right )}{{\rm log} \left ( R_{\rm min} / \rioni \right )} & (R < \rioni) \\
\ \ \ 2 & ({\rm otherwise}) \ ,
\end{cases}
\label{eq:alpha}
\end{eqnarray}
where $R_{\rm min}$ is the inner boundary of the disk model.
We note that the density profile is assumed so that the mass flux is set to 
$\mdot$ and $\minf$ at $R=R_{\rm min}$ and $\rioni$, respectively. 
It is worth noting that $\alpha=2$ even at $R<\rioni$ if the mass outflow rate is zero (see Eq. \ref{eq:rhocent}).

Given that the density structure is characterized with the two values of $\mdot$ and $\rioni$, 
we solve the radiation transfer equation for EUV photons within the disk, considering photoionization and dust absorption.
Note that we do not take into account diffusive EUV photons produced by radiative recombination of hydrogen.
The number flux of ionizing photons $Q(R)$ penetrating into the disk through the mid-plane is calculated by solving the equation of photon-number conservation
\begin{eqnarray}
\frac{{\rm d}\phi}{{\rm d} R} = - f(R) \phi - g(R) \ ,
\label{eq:ph_cons}
\end{eqnarray}
where $\phi (R)= Q/Q_0$ is the normalized photon flux and $Q_0$ is the unattenuated photon flux from the emission region, 
i.e., $\phi (R_{\rm min})=1$ is set.
The functions of $f(R)$ and $g(R)$ are given by 
\begin{eqnarray}
f(R) = \kappa_{\rm d, UV} \rho,
\label{eq:f}
\end{eqnarray}
\begin{eqnarray}
g(R) = \frac{4 \pi R^2 \alpha_{\rm B}}{Q_0} \left(\frac{\rho}{m_{\rm p}}\right)^2 \left (\frac{H_{\rm in}}{R_{\rm min}} \right ) \ .
\label{eq:g}
\end{eqnarray}
Here, $\alpha_{\rm B}$ is the case B recombination rate (the temperature is set at $T= 7 \times 10^4$ K)
and $H_{\rm in}$ is the disk height at $R=R_{\rm min}$.
The first and second term in the right-hand-side of Eq. (\ref{eq:ph_cons}) represent the effect of dust absorption
and radiative recombination of hydrogen.
This differential equation has an analytical solution of 
\begin{eqnarray}
\phi(R) = e^{-\tau(R)} \left \{ 1 - \int^R_{R_{\rm min}} g(R') e^{\tau(R')} {\rm d} R' \right \} \ ,
\label{eq:phi}
\end{eqnarray}
\begin{eqnarray}
\tau(R) = \int^R_{R_{\rm min}} f(R') {\rm d} R'  \ ,
\label{eq:tau}
\end{eqnarray}
where $\tau(R)$ corresponds to the optical depth for absorption of UV photons by dust.

In order to solve the radiative transfer equation, we set the photon number flux injected from the inner boundary.
For anisotropic radiation set by Eq. (\ref{eq:aniso_flux}), the ionizing photon number flux within the disk height is estimated as
\begin{eqnarray}
Q_0 &=& 3 \int^\infty_{\nu_{\rm T}} \frac{L_\nu}{h \nu} {\rm d} \nu \int^{\pi/2+\theta_{\rm d}}_{\pi/2-\theta_{\rm d}} {\rm cos}^2 \theta {\rm sin} \theta {\rm d} \theta \nonumber  \\
&=& \frac{L}{h \nu_{\rm T}} {\rm sin}^3 \theta_{\rm d} \simeq \frac{L}{h \nu_{\rm T}} \left ( \frac{H_{\rm in}}{R_{\rm min}} \right )^3 \ ,
\label{eq:photon}
\end{eqnarray}
where $\theta_{\rm d} = \tan^{-1}(H_{\rm in}/R_{\rm min})$ and $h \nu_T =$ 13.6 eV.
The choice of the inner boundary $R_{\rm min}$ seems somewhat arbitrary.
In this work, we adopt the dust sublimation radius defined by
\begin{equation}
\begin{split}
\rsb &=  \sqrt{\frac{L}{4\pi \sigma_{\rm SB} T_{\rm d}^4}} \\
&\sim 4.8 \times 10^3~{\rm AU}~\left( \frac{L}{\ledd}  \right)^{1/2} \left( \frac{\mbh}{10^4~\msun} \right)^{1/2} \left( \frac{T_{\rm d}}{1000~{\rm K}} \right)^{-2}  \ ,
\label{eq:rsb}
\end{split}
\end{equation}
where $\sigma_{\rm SB}$ is the Stefan-Boltzmann constant.
This choice is justified because at $R<\rsb$, EUV radiation is not absorbed by dust and photoionization substantially dominates radiative recombination; namely $d\phi (R)/dR \simeq 0$ is a good approximation.
By solving those non-linear equations numerically, the size of the HII region is calculated so that $\phi(\rioni)=0$,
and the updated value of $\rioni$ is used to set the density profile in Eq. (\ref{eq:alpha}).
As a result of the iterative calculations, the size of the HII region is numerically expressed as a function of the mass accretion rate;
$\rioni = \mathcal{F}(\mdot)$.

Finally, we derive the equilibrium solution of the mass accretion rate, for which the mass conservation 
($\mdot + \mout - \minf = 0$) is satisfied.
We approximately estimate the mass outflow rate from the disk surface 
due to photoevaporation as
\begin{eqnarray}
\mout =  
{\rm min} \left \{ \minf, \ 4 \pi c_{\rm s,HII} \int^{\rioni}_{R_{\rm B, HII}} \rho(R) R {\rm d} R \right \},
\label{eq:mout}
\end{eqnarray}
at $\rioni > R_{\rm B, HII}$\footnote{
Since $\rsb<R_{\rm B, HII}$ is basically satisfied, 
the photoevaporative mass-loss rate from the interior of $\rsb$ is negligible.
Thus, we can approximate the mass inflow rate at $R=\rsb$ as the mass accretion rate onto the central BH.}
, and otherwise $\mout =0$.
Here, $R_{\rm B,HII} = 2 \times 10^4~{\rm AU}$ and $c_{\rm s,HII} = 20~{\rm km~s^{-1}}$ are the Bondi radius and 
the sound speed for ionized gas with $T = 7 \times 10^4$ K, respectively.
We note that the total mass outflow rate within $r\leq \rioni$ as a representative value 
instead of the outflow rate of $\mout(>R)$.
Using the relation of $\rioni = \mathcal{F}(\mdot)$, the mass outflow rate is given by a function of $\mdot$,
denoting $\mout = \mathcal{G}(\mdot)$.
Therefore, the equilibrium value of $\mdot$ is obtained from the mass conservation law of
$\mdot + \mathcal{G}(\mdot) = \minf$.

In Figure \ref{fig:rst}, we present the equilibrium values of $\rioni/R_{\rm B,HII}$ (top) and 
$\mdot/\medd$ (bottom) as a function of $\fin$ and $Z$.
In the left column, the BH mass and disk sound speed are set to $\mbh = 10^4~\msun$ and 
$c_{\rm s} = 2~\rm km~s^{-1}$, corresponding to the situation seen in our RHD simulations.
With larger values of $\fin$ and $Z(\gtrsim 10^{-2}~\zsun)$, the size of the ionized region shrinks 
relative to the Bondi radius owing to efficient recombination and EUV absorption by dust.
Since photoevaporation ceases suddenly as the ratio of $\rioni/R_{\rm B,HII}$ 
becomes below unity (above the solid curve in each panel), most of the injected mass can 
feed the central BH at super-Eddington accretion rates without significant mass loss.
Otherwise, the mass accretion rate is strongly suppressed and limited below the Eddington rate 
because of mass loss led by photoevaporation (below the solid curve in each panel).
We note that the BH feeding rates obtained from the RHD simulations (filled circles) are 
nicely explained with this semi-analytical model.
In particular, we successfully demonstrate that the accretion disk with $Z = 10^{-3}~\zsun$ and $\fin = 100$ (Z-3F2 model) 
become optically thin to ionizing radiation and lose a larger amount of its mass owing to photoevaporation,
compared to the higher-metallicity case with $Z = 10^{-2}~\zsun$ (Z-2F2), as shown in Figures \ref{fig:mdot_summary} and \ref{fig:den_hz}.

In the middle panels of Figure \ref{fig:rst}, we present the case with a higher sound speed of 
$c_{\rm s} = 6~\rm km~s^{-1}$.
This corresponds to the higher-resolution cases (Z-2F2hr+ne and Z-2F3hr+ne), in which 
vertical gas motions dynamically heat the accretion flow.
In this case, the HII region expands substantially because the gas density in the (dynamically) 
hotter disk decreases (see Eq. \ref{eq:rhocent}).
As a result, even higher values of $\fin$ and $Z$ are required to sustain high BH accretion rates.

Moreover, to see the effect of BH mass growth, we demonstrate the case with a higher BH mass 
of $\mbh = 10^5~\msun$ in the semi-analytical model as shown in the right panels of Figure \ref{fig:rst} 
(although the BH mass is fixed to $\mbh = 10^4~\msun$ throughout our RHD simulation).
In this case, the conditions required for rapid BH accretion are relatively moderate; 
namely, the solid curve moves to the lower left.
This is because the larger-mass BH captures photoionized gas more effectively and 
thus a denser and optically-thick accretion disk forms.

In conclusion, the conditions required to avoid mass loss owing to photoevaporation 
(see solid curves in Figure \ref{fig:rst}) are approximately expressed as  
\begin{equation}
F_{\rm in} > 1000~
\left(\frac{M_{\rm BH}}{10^4~\msun}\right)^{-1}
\left(\frac{c_{\rm s}}{10~{\rm km~s}^{-1}}\right) 
\left ( 1 + \frac{Z}{10^{-2}~\zsun} \right )^{-1} \ .
\label{eq:Fcondi}
\end{equation}
Therefore, the super-Eddington condition ($\fin>1$) combined with Eq. (\ref{eq:Fcondi}) 
is expressed as
\begin{equation}
M_{\rm BH} < 8\times 10^6~\left(\frac{T}{10^4~{\rm K}}\right)^{1/2}
\left (1+\frac{Z}{10^{-2}~\zsun} \right )^{-1}~\msun,
\label{eq:Mcondi}
\end{equation}
where $T$ is the gas temperature in the disk.
We note that this criterion for BH mass is essentially equivalent to that obtained from
1D spherically symmetric RHD simulations by \cite{Inayoshi2016}, 
where the critical BH mass (see their Eq. 36) is derived from $R_{\rm B}<R_{\rm HII}$ 
by assuming a gas density distribution of $\rho(r) \propto r^{-2}$.
The difference of the critical mass comes from a geometrical effect between disk-like accretion
and spherical accretion.


\section{Possible sites for super-Eddington growth of seed BHs} \label{sec:possible}

In this section, we argue where and when super-Eddington mass growth of seed BHs
takes place in high-$z$ protogalaxies.
The critical conditions of Eq. (\ref{eq:Fcondi}) and $\fin > 1$ are rewritten as
\begin{eqnarray}
\minf > {\rm max} \left (\mcrit,~\medd \right ) \ ,
\label{eq:condi}
\end{eqnarray}
where 
\begin{equation}
\begin{split}
\mcrit \equiv 2.2 &\times 10^{-1}~\msun ~{\rm yr}^{-1} \\
&\left ( 1 + \frac{Z}{10^{-2}~\zsun} \right )^{-1} \left(\frac{c_{\rm s}}{10~{\rm km~s}^{-1}}\right) \ .
\label{eq:mdot_crit}
\end{split}
\end{equation}
Note that the condition for $\mcrit > \medd$ is equivalent to that in Eq. (\ref{eq:Mcondi}).
In what follows, assuming a simple galaxy evolution model, we evaluate the typical values of both $Z$ 
and $\dot{M}_{\rm in}$ that depend on the properties of the host galaxies and their assembly histories.

\begin{figure}
\begin{center}
\includegraphics[width=8.5cm]{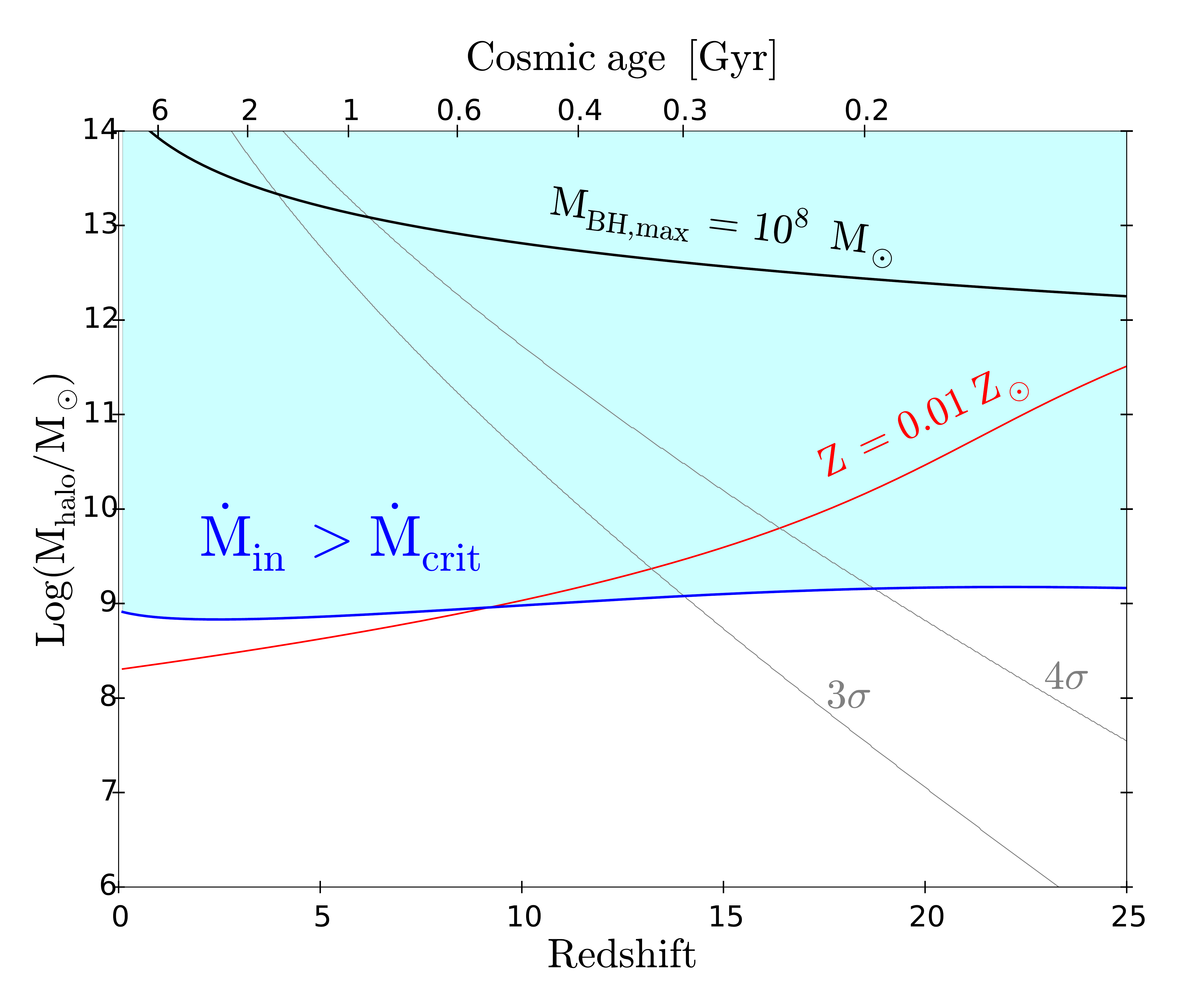}
\end{center}
\caption{Summary of the analytical argument in \S\ref{sec:possible}.
In the blue shaded area, the critical conditions required for super-Eddington accretion onto seed BHs are satisfied;
$\minf > \mcrit$ for $c_{\rm s} = 10~\rm km~s^{-1}$ (Eqs. \ref{eq:mdot_crit} and \ref{eq:minf2}).
The red and black curves show the halo mass where the metallicity reaches $Z = 0.01~\zsun$ (Eq. \ref{eq:mvzr}) and 
the efficient growth of BHs terminates at $M_{\rm BH,max} = 10^8~\msun$ (Eq. \ref{eq:mbh_max}), respectively.
In DM halos formed in overdense regions with a mass variance of $3-4~\sigma$ (grey curves),
super-Eddington mass growth of seed BHs would take place at $z \lesssim 15-20$.
}
\label{fig:mhalo_z}
\end{figure}

Although we are aimed to explore the growth of seed BHs at $z>6$, chemical evolution of galaxies has not
been understood properly and their observations are still limited at $z<4$
\citep[e.g.,][]{Mannucci2010, Troncoso2014, Hunt2016, Onodera2016}.
Instead, we here adopt a chemical-enrichment model proposed by recent cosmological simulations 
of galaxy formation \citep{Sarmento2018}.
Their simulations successfully reproduce the statistical properties of young galaxies hosted in DM halos with masses of $M_{\rm h}\sim 10^{8-11}~\msun$, 
such as the rest-frame UV luminosity functions observed at $z \geq 7$, and also predict
a halo-mass and metallicity relation at $z = 7-15$.
The $M_{\rm h}-Z$ relation can be fitted with a polynomial function of 
\begin{equation}
{\rm log}(Z/Z_\odot) = - 0.5 - 0.085 \ z + 0.48 \ m - 0.13 \ m^2  + 0.058 \ m^3 \ ,
\label{eq:mvzr}
\end{equation}
where $m \equiv {\rm log}(M_{\rm h}/10^{10}~M_\odot)$.
We note that the metallicity $Z$ in Eq. (\ref{eq:mvzr}) represents the mean value averaged over a galaxy, 
neglecting inhomogeneous chemical-enrichment in galaxies, although enriched gas tends to be concentrated to 
the inner regions of galaxies \citep[e.g.,][]{Vila-Costas1992, Luck2011, Sanchez2014, Toyouchi2014, 
Tissera2016, Tissera2019, Grand2019}.
Therefore, our model takes a lower limit of gas metallicity in the nuclear regions of galaxies.
The red curve in Figure \ref{fig:mhalo_z} shows the mass of DM halos where the averaged metallicity is 
$Z = 0.01~\zsun$ at each redshift $z$, estimated from Eq. (\ref{eq:mvzr}),
above which EUV attenuation by dust grains affects the critical conditions for BH rapid growth.
For instance, in a massive DM halo with $\mh \sim 10^9~\msun$, such metal-enriched regions form by $z \simeq 13$.

Next, we estimate the mass injection rate from galactic disk scales onto nuclear regions,
through a warm circum-nuclear disk with gas temperature of $T = 10^4$ K.
In the protogalactic nuclei, the gaseous disks tend to be gravitationally unstable \citep[e.g.,][]{Oh2002}
and thus the structure adjusts so that the Toomre's Q parameter 
is close to unity
\begin{eqnarray}
Q \equiv \frac{c_{\rm s,d} V_{\rm rot}}{\pi G \Sigma R} \simeq 1 , \
\label{eq:Q}
\end{eqnarray}
where $c_{\rm s,d}$ is the sound speed within the disk, and $V_{\rm rot}$ is the disk rotational velocity.
Motivated by the observations of star-forming disk galaxies \citep[e.g.,][]{Begeman1989, Swaters2000}, 
we assume a flat rotation curve, indicating that the rotational velocity is approximated as
the circular velocity of the DM halo ($V_{\rm rot} \simeq V_{\rm c,h}$).
%
With these assumptions, the mass accretion rate through the disk is expressed as
\begin{eqnarray}
\minf = 2 \pi R \Sigma v_{r} \simeq \frac{2  \mach c^2_{\rm s,d} V_{\rm vir} }{GQ}, \
\label{eq:minf2}
\end{eqnarray}
where $v_{r}$ is the radial velocity, and $\mach \equiv v_{r}/c_{\rm s,d}$ is the radial mach number.
Similarly to the semi-analytical star forming disk model described by \cite{Thompson2005}, 
angular momentum transport in the disk is assumed to be induced by axisymmetric spiral structures
and $\mach = 0.1$ is set based on a phenomenological prescription to describe this process
($\mach \lesssim 0.2$; see \citealt{Goodman2003}).
Since the mass inflow rate $\minf$ depends only on the properties of the DM halo, 
for the cases with $\mcrit > \medd$ or equivalently Eq. (\ref{eq:condi}),
the critical condition of $\minf > \mcrit$ is independent of the BH mass
as shown by the blue solid curve in Figure \ref{fig:mhalo_z}.
The resulting condition is given by $M_{\rm h}>10^9~\msun $ almost independently of $z$.
This suggests that DM halos formed in overdense regions with a mass variance of 3-4$\sigma$ 
become possible sites where super-Eddington mass growth of seed BHs would be led
during $z \sim$ 15-20.
Such massive halos are heavier than the mass of ``typical" direct-collapse BH forming halos 
\citep[see ][]{Volonteri2012, Haiman2013, Inayoshi2019}. 
These facts imply that even rarer populations of seed BHs could undergo efficient mass growth 
immediately after their formation, as pointed out in \cite{Inayoshi2019} \citep[see also][]{Valiante2016}.

When a seed BH is embedded in the center of a massive DM halo with $M_{\rm h} \gtrsim 10^9~\msun$,
the seed can undergo super-Eddington growth in mass regardless of their initial mass.
However, as the BH mass increases and $\mcrit < \medd$ is satisfied, the critical condition is given by 
$\minf >  \medd$, requiring an upper limit of the  BH mass
\begin{align}
 M_{\rm BH,max} & \simeq 1.1 \times 10^8~\msun~\left( \frac{\mh}{10^{13}~\msun} \right )^{1/3} \left( \frac{1+z}{10} \right )^{1/2} \nonumber \\
 & \simeq 7.9 \times 10^7~\msun \left(\frac{T_{\rm vir}}{10^7~{\rm K}}\right)^{1/2}.
\label{eq:mbh_max}
\end{align}
This also provides a condition for BH mass, above which rapid growing phases of seed BHs terminate 
due to the lack of mass reservoir in their host galaxies (see black solid curve in Figure \ref{fig:mhalo_z}).
This argument for seed BH growth is consistent with a scenario proposed by \cite{Inayoshi2016},
where spherically-symmetric rapid mass accretion is studied.
Their upper mass limit is estimated as $M_{\rm BH} \lesssim 1.4\times 10^8~\msun~ (T_{\rm vir}/10^7~{\rm K})$,
which differs from Eq. (\ref{eq:mbh_max}) by a factor of $\simeq 2$ and has a stronger dependence on 
$T_{\rm vir}$ due to different accretion geometry.

We note that super-Eddington accretion does not last eternally since the BH mass growth 
leads to $\medd (\propto M_{\rm BH}) > \minf$.
The efficient BH growth terminates when the BH mass becomes as high as $\sim 10^8~\msun$ 
even in massive halos associated with a mass variance of $3-4\sigma$.
This seems consistent with the existence of sub-Eddington, low-luminous quasars with 
$\mbh \gtrsim 10^8~\msun$ at $z \gtrsim 6$ \citep[e.g.,][]{Matsuoka2019, Onoue2019},
but might fail to explain the existence of brighter quasars with $L/\ledd \sim 1$ 
\citep[e.g.,][]{Willott2010b, Mortlock2011, Wu2015, Banados2018, Yang2020}.
Possibly, rapid accretion onto such SMBHs would be induced by a large amount of gas injection 
into nuclear regions associated with galaxy-galaxy major mergers, 
as seen in cosmological hydrodynamical simulations \citep[e.g.,][]{Hopkins2010}. 
Therefore, our argument that focuses on steady mass transport through marginally stable disks would 
give a conservative estimate of $\fin$.
To explore the nature of non-steady violent mass accretion consistent with the outer boundary conditions
set by large-scale cosmological simulations is left for future investigations.

\section{Discussion and caveats} \label{sec:effect}


\subsection{Stellar feedback} \label{sec:stellar}

Intense UV radiation and energetic supernovae associated with massive star formation 
heat the interstellar medium and potentially induce mass loss from galactic disks \citep[e.g.,][]{Hopkins2011, Hopkins2012, Li2015, Li2017, Kim2018}.
Cosmological hydrodynamical simulations, which explore the coevolutionary process of galaxies and SMBHs, 
have shown that stellar feedback substantially regulates the mass budget in the nuclear regions and suppresses 
mass feeding to the central BHs \citep[e.g.,][]{Dubois2015, Habouzit2017, Angles2017, Angles2020, Catmabacak2020}.
\cite{Latif2018} also demonstrated that the combination between stellar and AGN feedback prevents seed BHs
from growing to $\mbh \sim 10^9~\msun$ by $z \sim 6$.
In contrast, \cite{Di_Matteo2017}, where a quite huge simulation box of $(400~h^{-1}~\rm Mpc)^3$ is adopted, 
found that stellar feedback is not strong enough to quench the growth of relatively massive seeds in the early epoch.
They also suggested that the formation of SMBHs depends on tidal field exerted to the host halo.
In fact, in lower-spin halos, efficient gas inflows via cold streams can directly feed the nuclear region without 
forming large stellar disks that launch galactic outflows.
The accretion efficiency under more realistic mass injection from a star forming galactic disk 
will be addressed in our future work.

Active star formations (at least vigorous disk fragmentation) can still take place in the nuclear regions 
as shown in our RHD simulations with a higher resolution (Z-2F2hr+ne model). 
The mass of clumps is as massive as $M_{c} \sim 100~\msun$ and those clumps are located at $R \sim 2$-$3~\rcent$,
where the density of the surrounding gas is $\Sigma \sim 10^{-8}~\msun~\rm AU^{-2}$.
Therefore, the clumps accrete from the disk at a rate of
\begin{eqnarray}
\dot{M}_{\rm c} = \frac{3}{2} \Sigma \Omega (f_{\rm H} R_{\rm H})^2 \ ,
\label{eq:mdotc}
\end{eqnarray}
where the Hill radius is $R_{\rm H} \equiv R (M_{\rm c}/3\mbh)^{1/3}$, and $f_{\rm H} \sim O(1)$ 
\citep[e.g.,][]{Goodman2004, Inayoshi2014b}.
Adopting the clump properties seen in the Z-2F2hr+ne model, we estimate the rate as 
\begin{equation}
\begin{split}
\dot{M}_{\rm c} \simeq 2.3 \times &10^{-4}~{\rm \msun~yr^{-1}} \\ 
&\left ( \frac{f_{\rm H}}{1.5} \right )^{2} \left ( \frac{R}{2.5 \times 10^5~\rm AU} \right )^{2}  \left ( \frac{M_{\rm c}}{100~\msun} \right )^{2/3}  \ .
\label{eq:mdotc2}
\end{split}
\end{equation}
Assuming that all the gas in a clump accretes to a single star, the growth rate of a newly-born protostar is given by 
$\dot{M}_{\rm c}$.
The accreting protostar begins to contract by loosing energy via radiative diffusion and evolves to a main-sequence star
in a Kelvin Helmholtz (KH) timescale of
\begin{equation}
\begin{split}
\tau_{\rm KH} &\sim \frac{M_{\rm c}}{\dot{M}_{\rm c}} \\
&\simeq 0.4~{\rm Myr}~\left ( \frac{M_{\rm c}}{100~\msun} \right ) \left ( \frac{\dot{M}_{\rm c}}{2.3 \times 10^{-4}~\msun \rm yr^{-1}} \right )^{-1}  \ ,
\label{eq:tkh}
\end{split}
\end{equation}
where we assume that KH contraction balances with energy input by protostellar accretion.
Since this timescale is shorter than the orbital time scale $\tau_{\rm orb} \equiv 2 \pi / \Omega \sim$ 1 Myr,
which is comparable to the clump migration timescale in a marginally stable disk \citep[e.g.,][]{Inayoshi2014b},
massive main-sequence stars form within the nuclear accretion disks and contribute to stellar radiative feedback.

Contrary to the negative stellar feedback that operate in larger-scales as discussed above,
starbursts events in the nuclear disk regions could potentially enhance mass transport through the disk
due to strong turbulence excited by SNe in a star-forming CNDs \citep{Wada2002, Kawakatu2008, Wutschik2013}. 
Indeed, a positive correlation between the AGN and star-formation activities in the nuclear region has been 
observed \citep[e.g.,][]{Diamond-Stanic2012, Esquej2014} and a state transition from quiescent phases to AGNs
would be triggered by nuclear starbursts \citep{Inayoshi2020}.
According to a semi-analytical model \citep{Kawakatu2009}, the BH feeding rate peaks when the gas supplying rate 
to the disk region is comparable to the gas consumption rate due to star formation,
requiring a high injection rate of $\minf \sim 10^3~\msun \rm yr^{-1}$ over $\sim 100$ Myr 
to explain the existence of SMBHs at $z \sim 6$.
To reveal whether star formation in the nuclear regions provides negative or positive effects on the mass growth 
of seed BHs is left for our future studies.

\subsection{BH radiative and mechanical feedback} \label{sec:wind}

In our RHD simulations, we treat the mass inflow rate at the inner boundary to be the BH accretion rate,
assuming the properties of radiative output (e.g., luminosity, spectra, and anisotropy) from the unresolved small scales.
We here briefly discuss the effect of those assumption and other types of radiative/mechanical output from
the vicinity of the nuclear accreting BH.

First, we consider the IR radiative force caused by reemission from heated dust grain.
Although our simulations include this effect, no significant impacts on the accretion flow are found
because the disk surface density is not high enough for the gas to trap diffuse IR photons.
In fact, the optical depth to IR photons toward the disk vertical direction is estimated as
\begin{eqnarray}
\tau_{\rm IR} &=& \kappa_{\rm d,IR} \Sigma \nonumber \\
&\simeq& 0.2~\left(\frac{F_{\rm in}}{100}\right)
~\left(\frac{Z}{10^{-2}~\zsun}\right)
~\left(\frac{M_{\rm BH}}{10^4~\msun}\right)^{1/4} , 
\label{eq:tau_ir}
\end{eqnarray}
where the surface density is estimated at $\rsb$, assuming the disk to be a steady state,
$\Sigma = \fin \medd / (2\pi \rsb v_R)$. 
Therefore, unless $\fin > 1000$ and $Z > 0.1~\zsun$ are considered, the critical conditions for rapid accretion do not change with 
the optical thickness of the gaseous disk against IR photons, as discussed in \S\ref{sec:condition}.

Analytical arguments by \cite{Krolik2007} and \cite{Shi2008} concluded that 
a geometrically thick disk supported by IR radiation pressure forms around the dust sublimation radius.
Several numerical simulations have confirmed that such disk structure produces outflows driven by 
IR radiation pressure onto dust and does not feed the central BH at a high rate 
\citep[e.g.,][]{Dorodnitsyn2012,Chan2016}.
The strong IR radiation pressure within the disk height is due to higher opacity of hotter dust 
that is heated by isotropic radiation emitted from the nuclear BH.
In contrast, assuming that the radiation flux from the nuclear BH is highly collimated toward the poles,
dust grain in the disk region is kept cold and the IR radiation pressure is significantly reduced even for a high accretion rate of $\mdot \sim 0.8~\medd$ \citep{Namekata2016}.
As a result, the accretion disk becomes geometrically thinner than what the previous analytical studies predicted.
Even extending to higher values of $\fin$ and $Z$ in our case, the disk accretion dynamics is not 
affected by IR radiation pressure, as long as the emergent radiation flux is sufficiently anisotropic.
We note, however, that if intense UV radiation from massive stars formed in a dusty disk heats dust grain
near the equatorial region, an IR-radiation-pressure supported disk forms in the nuclear region
\cite[e.g.,][]{Thompson2005}. 

In addition to radiative feedback, mechanical feedback due to outflows and disk winds launched from the vicinity of the BH
would affect mass accretion in a CND region \citep[e.g.,][]{Fabian2012}, although we do not explicitly inject mechanical
momentum from the inner boundary in our simulations.
One possible mechanism to launch outflows is the line-driven wind model, where UV radiation emitted from the disk 
around $\sim 100-1000~R_{\rm Sch}$ accelerates moderately ionized metal gas yielding substantially high opacity 
via bound-bound transitions.
In the wind regions, the radiative force caused by various spectral lines boosts the acceleration efficiency 
by several orders of magnitude above the continuum radiation force exerted through electron scattering alone
\citep{Stevens1990, Proga2000, Proga2004,Nomura2020}.
Another possible channel is that a highly accreting BH with super-Eddington luminosity ($L\gg L_{\rm Edd}$)
exerts the radiation force through electron scattering in optically thick medium and produce strong outflows
\citep[e.g.,][]{Ohsuga2005, Ohsuga2011, Jiang2014, Yang2014, Yang2018, Sadowski2016}.
In both cases, a large fraction ($\gtrsim 50\%$) of the injected mass from larger scales is loaded 
into outflows collimated toward the polar regions.
Since the BH feeding rate is reduced due to mass loading to outflows and the radiative luminosity decreases,
the presence of disk winds rather promotes rapid accretion more efficiently \citep{Takeo2020}.
This fact suggests that we need to make a comprehensive model of BH accretion covering the outflow launching scale
and the Bondi scales, in order to better understand the BH growth mechanism.


\section{SUMMARY AND CONCLUSION} \label{sec:summary}

In this paper, we study rapid mass accretion onto IMBHs with $\mbh = 10^4~\msun$ 
embedded in massive self-gravitating, dusty nuclear accretion disks, performing the
first 3D RHD simulations focusing on the nuclear region of protogalaxies. 
Our simulations resolve the Bondi radius for hot ionized gas with a temperature of 
$T \sim 10^5$ K and can follow the launching process of outflows from the disk surface
owing to photoevaporation which suppresses the BH from accreting.

We here explore the dependence of mass accretion efficiency on the gas metallicity $Z$ and
mass injection rate from the outer galactic disk normalized by the Eddington vale $\fin \equiv \minf/\medd$.
For this purpose, we run several numerical models (e.g., Z2F2 and Z2F1 models) covering
a wide range of the relevant parameters of $\fin =$ 10-1000 and $Z = 10^{-3}$-$10^{-1}~\zsun$.
In all cases, the nuclear disk becomes gravitationally unstable and 
transports mass inward owing to angular momentum transport caused by global density spiral arms.
The central BH can be fed at rates exceeding the Eddington rate only when
the dusty disk becomes sufficiently optically thick to ionizing radiation.
In this case, a large fraction ($\gtrsim 40\%$) of the mass injection rate can feed the central BH.
The critical conditions are given by $\minf > \mcrit$, where 
\begin{equation}
\begin{split}
\mcrit \equiv 2.2\times & 10^{-1}~\msun ~{\rm yr}^{-1} \nonumber \\ 
& \left ( 1 + \frac{Z}{10^{-2}~\zsun} \right )^{-1} \left(\frac{c_{\rm s}}{10~{\rm km~s}^{-1}}\right) \ ,
\end{split}
\end{equation}
and $c_{\rm s}$ is the sound speed in the gaseous disk.
Otherwise, since the disk is not obscured enough to shield intense ionizing radiation by dust absoption,
mass outflows from the disk owing to photoevaporation limit the BH accretion rate to $\simeq 1-10\%$ 
of the mass injection rate from the outer boundary and thus strongly prevent the BH feeding.

With the cases where a higher numerical resolution is set and the equatorial-symmetric assumption 
is relaxed (model Z-2F2hr+ne and Z-2F3hr+ne), vigorous disk fragmentation reduces the disk surface density 
and dynamical heating by formed clumps makes the disk thickness higher.
As a result, the photoevaorative mass-loss rate rises and thus the critical injection rate increases.
However, the central BH can be fed at super-Eddington rates once $\minf > \mcrit$ is satisfied even if the disk becomes dynamically hot owing to clump formation.

Finally, we apply our results to the cosmological evolution of massive BHs via rapid mass accretion.
With a semi-analytical model, we find that super-Eddington accretion is allowed until the BH mass 
reaches $\mbh \sim 10^{7-8}~\msun$, depending on the properties of the host DM halo and 
metal-enrichment history.
In the assembly of protogalaxies, seed BHs that form in overdense regions with 
a mass variance of 3-4$\sigma$ at $z\sim 15-20$ are able to undergo short periods of their rapid growth 
and transits into the Eddington-limited growth phase afterwards to be SMBHs observed at $z>$ 6-7.

\acknowledgments

\clearpage

The authors would like to thank Masayuki Umemura, Tohru Nagao, Shingo Hirano, and Naoki Yoshida for fruitful discussions,
and Kazuyuki Sugimura and Riouhei Nakatani for their contribution to developing the numerical code. 
The numerical simulations were performed with the Cray XC50 at the Center for Computational Astrophysics (CfCA) of the National Astronomical Observatory of Japan and with High-performance Computing Plat- form of Peking University. 
This work is financially supported by the National Science Foundation of China (11721303, 11991052, 11950410493; KI), and the National Key R\&D Program of China (2016YFA0400702; KI), the Grants-in-Aid for Basic Research by the Ministry of Education, Science and Culture of Japan (17H06360: D.T., 16H05996, 17H01102, 19H01934: T.H.). 
R.K. acknowledges financial support via the Emmy Noether Research Group on Accretion Flows and Feedback in Realistic Models of Massive Star Formation funded by the German Research Foundation (DFG) under grant no. KU 2849/3-1 and KU 2849/3-2.

\bibliography{ref}{}
\bibliographystyle{aasjournal}

\end{document}